\pdfoutput=1
\documentclass[11pt,twoside,a4paper,cmspaper,final,collab]{cms-tdr}
\def\svnVersion{ee05af1}\def\svnDate{2022/01/28}\def\cmsCernNoTag{CERN-EP-2021-161}\def\cmsCernDate{\today}\def\cmsMessage{Published in the Journal of High Energy Physics as \href{http://dx.doi.org/10.1007/JHEP01(2022)188}{\doi{10.1007/JHEP01(2022)188}.}}
\begin{document}\cmsNoteHeader{SMP-20-010}

\newcommand{\jet}{\ensuremath{\mathrm{j}}}
\newcommand{\lambdakb}{\ensuremath{\lambda^{\kappa}_{\beta}}\xspace}
\newcommand{\multi}{\ensuremath{\lambda^{0}_{0}}\xspace}
\newcommand{\lha}{\ensuremath{\lambda^{1}_{0.5}}\xspace}
\newcommand{\ptd}{\ensuremath{\lambda^{2}_{0}}\xspace}
\newcommand{\pTD}{\ensuremath{(p_{\mathrm{T}}^{\mathrm{D}})^2}\xspace}
\newcommand{\width}{\ensuremath{\lambda^{1}_{1}}\xspace}
\newcommand{\thrust}{\ensuremath{\lambda^{1}_{2}}\xspace}
\newcommand{\CA}{\ensuremath{C_{\mathrm{A}}}\xspace}
\newcommand{\CF}{\ensuremath{C_{\mathrm{F}}}\xspace}
\newcommand{\Zjet}{\ensuremath{\PZ{+}\text{jet}}\xspace}
\newcommand{\mgpy}{MG{5}+\PYTHIA{8}}

\cmsNoteHeader{SMP-20-010}
\title{Study of quark and gluon jet substructure in \Zjet and dijet events from \texorpdfstring{$\Pp\Pp$}{pp} collisions}

\author*[inst1]{CMS Collaboration}

\date{\today}

\abstract{
Measurements of jet substructure describing the composition of quark- and gluon-initiated jets are presented.
Proton-proton ($\Pp\Pp$) collision data at $\sqrt{s}=13\TeV$ collected with the CMS detector are used, corresponding to an integrated luminosity of $35.9\fbinv$.
Generalized angularities are measured that characterize the jet substructure and distinguish quark- and gluon-initiated jets.
These observables are sensitive to the distributions of transverse momenta and angular distances within a jet.
The analysis is performed using a data sample of dijet events enriched in gluon-initiated jets, and, for the first time, a \Zjet event sample enriched in quark-initiated jets.
The observables are measured in bins of jet transverse momentum, and as a function of the jet radius parameter.
Each measurement is repeated applying a ``soft drop" grooming procedure that removes soft and large angle radiation from the jet.
Using these measurements, the ability of various models to describe jet substructure is assessed, showing a clear need for improvements in Monte Carlo generators.
}

\hypersetup{
pdfauthor={CMS Collaboration},
pdftitle={Study of quark and gluon jet substructure in Z+jet and dijet events from pp collisions},
pdfsubject={CMS},
pdfkeywords={CMS,  measurement, jet substructure, quarks, gluons, dijet, Z+jet}}

\maketitle

\section{Introduction}
\label{sec:intro}

Interactions between quarks and gluons, commonly referred to as partons, are governed by quantum chromodynamics (QCD).
When quarks or gluons are produced by collisions in particle colliders, collimated showers of hadrons are formed.
Theoretical calculations and experimental methods cluster such showers into jets using dedicated clustering algorithms~\cite{jetography} that correlate the kinematic properties of the jets with those of the participating partons.
Furthermore, the study of the distribution of particles within a jet, the jet substructure, can be analyzed to determine the likelihood that a jet originates either from a quark or from a gluon.
Heavy, Lorentz-boosted particles that decay to quarks and/or gluons, \eg top quarks or {\PW}, {\PZ} or Higgs bosons, have characteristic distributions of jet substructure observables.
Several algorithms have been developed to distinguish such jets from those originating from single partons~\cite{Asquith:2018igt, Larkoski:2017jix, Marzani:2019hun}.
These are used in numerous measurements and searches at the CERN LHC, and frequently the modelling of jet substructure contributes significantly to the systematic uncertainties~\cite{Asquith:2018igt}.
Many of them would benefit from a better understanding and simulation of jet substructure observables.

Although there is no unique way of defining whether a jet originates from a quark or gluon in QCD~\cite{Gras:2017jty}, at leading order (LO) in perturbation theory, one can label jets according to the parton initiating the particle shower.
Jets originating from single quarks, quark jets, and those arising from single gluons, gluon jets, are known to differ in their substructure.
At LO in perturbative QCD, the difference in the Casimir colour factor between quarks ($\CF = 4/3$) and gluons ($\CA = 3$) leads to a higher probability for gluons to radiate a soft gluon by a factor of  $\CA/\CF = 9/4$~\cite{Konishi:1979cb}.
Therefore, we expect gluon jets to contain, on average, more constituents, with a wider geometric spread in the detector~\cite{OPAL:1993uun,OPAL:1995ab,DELPHI:1995nzf,ALEPH:1998pel,ALEPH:1995oxo,Barate:1996fi}.
However, there are also a variety of other nonperturbative effects that determine the jet substructure, including hadronization and colour reconnection, which are typically described by phenomenological models in Monte Carlo (MC) event generators.
The description of jet fragmentation and substructure thus requires a combined understanding of both perturbative calculations and nonperturbative effects.
Although the description of jets initiated by quarks is well constrained by many previous measurements, the modelling of gluon jets is less well understood~\cite{Gras:2017jty, Asquith:2018igt}.

Previous measurements of observables sensitive to jet fragmentation at the LHC have been reported by the ATLAS~\cite{Aad:2019onw, Aaboud:2018uiu, Aaboud:2017tke, Aaboud:2017bzv, Aad:2016oit, Aad:2011sc, Aad:2011gn}, CMS~\cite{Sirunyan:2018ncy, Sirunyan:2017bsd, Chatrchyan:2014ava, Chatrchyan:2012mec, Chatrchyan:2012gw}, ALICE~\cite{Acharya:2019djg, Acharya:2019zup, Acharya:2018edi, Acharya:2018eat}, and LHCb~\cite{Aaij:2019ctd} Collaborations.
Jet substructure quantities for the separation of jets initiated by quarks, gluons, {\PW}, {\PZ}, and Higgs bosons, and top quarks have been measured by ATLAS~\cite{Aad:2020zcn, Aad:2019vyi, Aaboud:2019aii, Aad:2019wdr, Aaboud:2018ibj, Aaboud:2017qwh, Aad:2015lxa, Aad:2015cua, Aad:2013fba, Aad:2012meb, ATLAS:2012am}, CMS~\cite{Sirunyan:2018asm, Sirunyan:2018xdh, Sirunyan:2017yar, Sirunyan:2017tyr, Chatrchyan:2013vbb} and ALICE~\cite{ALICE:2021obz, ALICE:2021njq}.
The goal of this measurement is to systematically explore a set of observables in multiple dimensions of phase space.
As outlined in Ref.~\cite{Gras:2017jty}, multiple aspects of MC event simulation models of jet substructure can be tested with such a comprehensive set of measurements.

\begin{figure}[htb]
\centering
\includegraphics[width=0.3\textwidth]{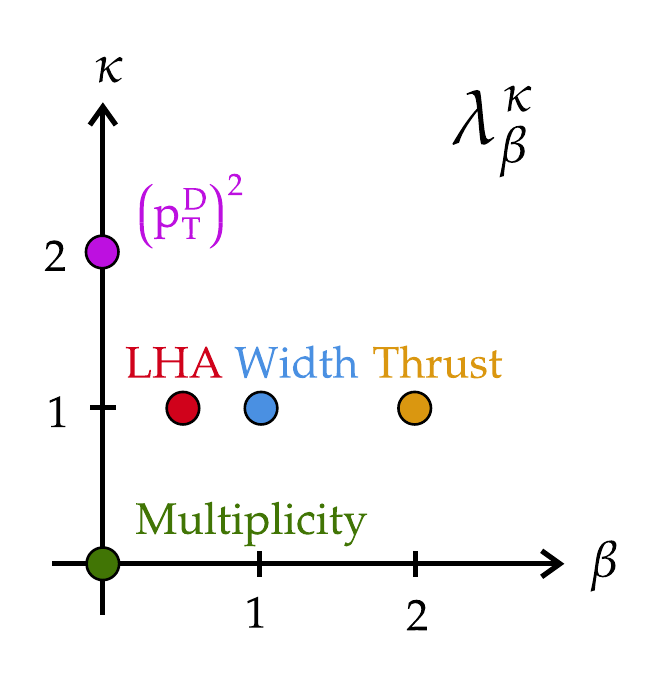}
\caption{The five generalized angularities \lambdakb used in this analysis, represented in the $(\kappa, \beta)$ plane. The Les Houches Angularity is denoted by LHA. Adapted from Ref.~\cite{Larkoski:2014pca}.}
\label{fig:lambdadiagram}
\end{figure}

In this paper, we report on a measurement of a set of five observables that can be used to distinguish between quark- and gluon-initiated jets, the generalized angularities \lambdakb~\cite{Larkoski:2014pca}, defined as:
\begin{linenomath}
\begin{equation}
\lambdakb = \sum_{i \in \text{jet}}z_{i}^{\kappa} \left(\frac{\Delta R_{i}}{R}\right)^{\beta},
\label{angularities}
\end{equation}
\end{linenomath}
where $z_{i}$ is the fractional transverse momentum carried by the $i$th jet constituent, $R$ is the jet radius parameter, and $\Delta R_{i}$ is the displacement of the constituent from the jet axis, defined as $\Delta R_{i} = \sqrt{\smash[b]{(\Delta y_{i})^{2} + (\Delta \phi_{i})^{2}}}$ where $\Delta y_{i}$ and $\Delta \phi_{i}$ are the separations in rapidity and azimuthal angle (in radians), respectively, between the jet axis and the ${i}$\textsuperscript{th} constituent.
The parameters $\kappa \geq 0$ and $\beta \geq 0$ control the momentum and angular contributions, respectively.
The five observables measured in this paper are: Les Houches Angularity (LHA) $= \lha$~\cite{Badger:2016bpw}; width $= \width$; thrust $= \thrust$~\cite{PhysRevLett.39.1587}; multiplicity $= \multi$; and $\pTD = \ptd$.
These are shown in Fig.~\ref{fig:lambdadiagram} as points in the $(\kappa, \beta)$ plane.
These quantities have previously been used by the CMS~\cite{CMS-PAS-JME-13-002,CMS-PAS-JME-16-003} and ATLAS~\cite{Aad:2014gea} Collaborations to discriminate between quark and gluon jets.
They have also been proposed as a tool to measure parton distribution functions~\cite{Caletti:2021ysv} and double-parton scattering~\cite{Kumar:2019twx} in $\PZ{+}\text{jets}$ events.

We measure these observables in quark-jet enriched \Zjet and gluon-jet enriched dijet samples.
In the \Zjet sample, we study a jet recoiling against a \PZ boson decaying to a pair of muons.
The measurement focuses on muons rather than electrons, since effects from bremsstrahlung are negligible for them.
Within the dijet event sample, we separately measure the distributions for the more central (smaller absolute rapidity $\abs{y}$) of the two jets, and the more forward (larger $\abs{y}$) of the two jets.
The sample of the more central jets is predicted to provide a higher purity of gluon jets~\cite{Aad:2016oit}.
Each quantity is measured as a function of jet transverse momentum \pt, with different jet radius parameters, including all particles and only charged particles, and with and without applying a \emph{soft drop} grooming procedure~\cite{Larkoski:2014wba} that removes soft and wide angle radiation from the jet, thereby making the jet substructure observables more resilient to effects from pileup, underlying event, and initial state radiation.
The detector level distributions are unfolded to particle level, \ie only considering final state particles after hadronization with proper lifetime $c\tau > 10\unit{mm}$, for each configuration and observable, and normalized to unity in each \pt bin.
The measurements are finally compared with predictions based on analytic resummation~\cite{Caletti:2021oor}.

After a brief description of the CMS detector in Section~\ref{sec:cms}, details about the data and simulated samples are given in Section~\ref{sec:samples}, followed by a description of the event selection in Section~\ref{sec:selection}.
The jet substructure observables are then more precisely introduced in Section~\ref{sec:observables}.
The unfolding procedure and uncertainties of the measurement are described in Section~\ref{sec:unfolding}.
Finally, results are presented and discussed in Section~\ref{sec:results}, and briefly summarized in Section~\ref{sec:conclusions}.

\section{The CMS detector and event reconstruction}
\label{sec:cms}

The central feature of the CMS apparatus is a superconducting solenoid of 6\unit{m} internal diameter, providing a magnetic field of 3.8\unit{T}. Within the solenoid volume are a silicon pixel and strip tracker, a lead tungstate crystal electromagnetic calorimeter (ECAL), and a brass and scintillator hadron calorimeter (HCAL), each composed of a barrel and two endcap sections. Forward calorimeters extend the pseudorapidity ($\eta$) coverage provided by the barrel and endcap detectors. Muons are detected in gas-ionization chambers embedded in the steel flux-return yoke outside the solenoid.

A more detailed description of the CMS detector, together with a definition of the coordinate system used and the relevant kinematic variables, can be found in Ref.~\cite{Chatrchyan:2008zzk}.

The silicon tracker measures charged particles within the range $\abs{\eta}<2.5$. It consists of 1440 silicon pixel and 15\,148 silicon strip detector modules. For particles of $1<\pt<10\GeV$ and $\abs{\eta}<1.4$, the track resolutions are typically 1.5\% in \pt and 25--90 (45--150)\mum in the transverse (longitudinal) impact parameter~\cite{TRK-11-001}.

The candidate vertex with the largest value of summed physics-object $\pt^2$ is the primary proton-proton ($\Pp\Pp$) interaction vertex. The physics objects are jets, clustered using the anti-\kt jet finding algorithm~\cite{Cacciari:2008gp,Cacciari:2011ma} with the tracks assigned to a candidate vertex as inputs, and the associated missing transverse momentum, which is the negative vector sum of the \pt of those jets.

Muons are measured in the pseudorapidity range $\abs{\eta}<2.4$, with detection planes made using three technologies: drift tubes, cathode strip chambers, and resistive plate chambers. The single muon trigger efficiency exceeds 90\% over the full $\eta$ range, and the efficiency to reconstruct and identify muons is greater than 96\%. Matching muons to tracks measured in the silicon tracker results in a relative transverse momentum resolution, for muons with \pt up to 100\GeV, of 1\% in the barrel and 3\% in the endcaps. The \pt resolution in the barrel is better than 7\% for muons with \pt up to 1\TeV~\cite{Sirunyan:2018}.

The particle-flow algorithm~\cite{CMS-PRF-14-001} is used to reconstruct and identify each individual particle in an event, with an optimized combination of information from the various elements of the CMS detector. The energy of photons is obtained from the ECAL measurement. The energy of electrons is determined from a combination of the electron momentum at the primary interaction vertex as determined by the tracker, the energy of the corresponding ECAL cluster, and the energy sum of all bremsstrahlung photons spatially compatible with originating from the electron track. The energy of muons is obtained from the curvature of the corresponding track. The energy of charged hadrons is determined from a combination of their momentum measured in the tracker and the matching ECAL and HCAL energy deposits, corrected for the response function of the calorimeters to hadronic showers. Finally, the energy of neutral hadrons is obtained from the corrected ECAL and HCAL energies.

For each event, hadronic jets are clustered from these reconstructed particles using the infrared and collinear (IRC) safe anti-\kt algorithm~\cite{Cacciari:2008gp, Cacciari:2011ma} with distance parameters of 0.4 and 0.8, referred to as AK4 and AK8 jets in the following. Jet momentum is determined as the vectorial sum of all particle momenta in the jet. It is found from simulation to be, on average, within 5 to 10\% of the true momentum over the whole \pt spectrum and detector acceptance.

Additional \Pp{}\Pp{} interactions within the same or nearby bunch crossings (pileup) can contribute additional tracks and calorimetric energy depositions to the jet momentum.
The pileup per particle identification algorithm~\cite{Bertolini:2014bba,Sirunyan:2020foa} is used to mitigate the effect of pileup at the level of reconstructed particles, making use of the local momentum distribution, event pileup properties and tracking information.
For each neutral particle, a momentum distribution variable $\log\sum{\pt^2/\Delta R^2}$ is computed using the surrounding charged particles compatible with the primary vertex within the tracker acceptance ($\abs{\eta}<2.5$), and using both charged and neutral particles in the region outside of the tracker coverage, where $\pt$ and $\Delta R$ correspond to the transverse momenta and distances of the surrounding particles in the $\eta$-$\phi$ plane w.r.t. the particle direction.
The probability to originate from the primary interaction vertex is deduced by comparing the momentum distribution variable to its event median characterizing the expected value for particles from pileup vertices. The momenta of the neutral particles are rescaled according to their probability, superseding the need for jet-based pileup corrections~\cite{Sirunyan:2020foa}.
Charged particles identified to be originating from pileup vertices are discarded.

Jet energy corrections are derived from simulation to adjust the measured response of jets to that of particle level jets on average. In situ measurements of the momentum balance in dijet, $\text{photon+jet}$, \Zjet, and multijet events are used to account for any residual differences in jet energy scale in experimental data and simulation~\cite{Khachatryan:2016kdb}.
The jet energy resolution amounts typically to 15\% at 10\GeV, 8\% at 100\GeV, and 4\% at 1\TeV.
Additional selection criteria are applied to each jet to remove jets potentially dominated by anomalous contributions from various subdetector components or reconstruction failures~\cite{Sirunyan:2020foa}.

Events of interest are selected using a two-tiered trigger system~\cite{Khachatryan:2016bia}. The first level, composed of custom hardware processors, uses information from the calorimeters and muon detectors to select events at a rate of around 100\unit{kHz} within a fixed latency of about 4\mus~\cite{Sirunyan:2020zal}. The second level, known as the high-level trigger, consists of a farm of processors running a version of the full event reconstruction software optimized for fast processing, and reduces the event rate to around 1\unit{kHz} before data storage.

\section{Data and simulated samples}
\label{sec:samples}

The analysis is based on $\Pp\Pp$ collision data collected by the CMS experiment in 2016 at a centre-of-mass energy of $\sqrt{s}=13\TeV$, corresponding to an integrated luminosity of $35.9\fbinv$ and with an average of 23 interactions per bunch crossing.
Events are selected by the trigger system using multiple sets of trigger paths.
The \Zjet candidate events are collected with a trigger requiring at least one isolated muon with $\pt>24\GeV$ and $\abs{\eta}<2.4$; the efficiency to select an event with at least one muon exceeds 90\%~\cite{Sirunyan:2021zrd}.
Dijet events with high-\pt jets are collected with triggers requiring at least one AK4 or AK8 jet above a certain threshold in \pt with an efficiency exceeding 99\%.
The triggers with $\pt<450\GeV$ are ``prescaled'', \ie only a fraction of the collision events passing the trigger requirements are accepted, which results in a lower integrated luminosity for these triggers.
Dijet events with jet $\pt<40\GeV$ are collected with a zero bias trigger that randomly selects a fraction of the collision events.
Table~\ref{tab:triggers} summarizes the zero bias and jet triggers with their integrated luminosity and the offline \pt bin thresholds for which the triggers are used.

\begin{table}[htb]
\centering
\topcaption{Summary of zero bias and jet triggers used in the analysis for the dijet region. For each trigger, the integrated luminosity and number of events collected by it are given. The offline \pt bin threshold(s) indicate the lower edge of the \pt bin(s) measured with data from a given trigger.}
\begin{tabular}{cccc}
Online \pt threshold & Integrated luminosity & Offline \pt bin threshold(s) & Number of events \\
(\GeVns{}) & (\fbinv) & (\GeVns{}) & \\
\hline
0 (zero bias) & $2.90 \times 10^{-5}$ & 50 & 309 512 \\
40 & $2.64 \times 10^{-4}$ & 65  & 446 183 \\
60 & $7.19 \times 10^{-4}$ & 88 & 387 849 \\
80 & $2.73 \times 10^{-3}$ & 120, 150 & 757 778 \\
140 & $2.39 \times 10^{-2}$ & 186 & 1 014 702 \\
200 & $1.03 \times 10^{-1}$ & 254 & 841 507 \\
260 & $5.88 \times 10^{-1}$ & 326 & 1 821 727 \\
320 & 1.75 & 408 & 1 918 614 \\
400 & 5.14 & 481 & 1 960 323 \\
450 & 35.9 & 614, 800, 1000 & 6 421 782 \\
\end{tabular}
\label{tab:triggers}
\end{table}

The processes of interest in this analysis are the \PZ boson production in association with jets, and events uniquely composed of jets produced through the strong interaction (QCD multijet production).
Their simulation is performed with multiple combinations of MC event generators to estimate the detector response and systematic uncertainties.
The \PZ boson production in association with jets is simulated at LO with the \MGvATNLO v2.2.2~\cite{Alwall:2014hca} generator with up to four outgoing partons in the matrix element calculations, and fragmented with \PYTHIA v8.212~\cite{Sjostrand:2014zea}.
\PYTHIA{8} implements a dipole shower ordered in \pt and the hadronization of quarks and gluons into final hadrons is described by the Lund string model~\cite{Andersson:1983ia,Sjostrand:1984iu}.
Double counting of jets from the matrix element calculation and parton shower is eliminated using the MLM merging scheme~\cite{Alwall:2007fs}.
This sample is hereafter referred to by {\mgpy}.
The \PYTHIA{8} parameters for the underlying event are set according to the CUETP8M1 tune~\cite{Skands:2014pea,Khachatryan:2015pea}.
A second \Zjet sample is generated at LO with \HERWIGpp v2.7.1~\cite{Bahr:2008pv,Bellm:2013hwb} in standalone mode with the UE-EE-5C underlying event tune~\cite{Seymour:2013qka}
to assess systematic uncertainties related to the modelling of the parton showering and hadronization.
In \HERWIGpp, the parton shower follows angular-ordered radiation~\cite{Gieseke:2003rz}, and the hadronization is described by the cluster model~\cite{Webber:1983if}.
This sample has one outgoing parton in the matrix element calculations, recoiling against the \PZ boson.
The \MGvATNLO and \HERWIGpp samples are normalized to the differential cross sections at next-to-LO (NLO) in electroweak and next-to-NLO (NNLO) in strong interactions~\cite{Lindert:2017olm}.

The QCD multijet production is simulated separately with two different generator configurations: \MGvATNLO combined with \PYTHIA{8}, and \HERWIGpp in a standalone mode.
In the former sample, again referred to by {\mgpy}, up to four outgoing partons are allowed in the matrix element calculations using the MLM jet merging scheme.
In the standalone \HERWIGpp sample, the matrix element has two outgoing partons, and the same underlying event tune is used as in the \HERWIGpp \Zjet sample.
The LO NNPDF 3.0~\cite{Ball:2014uwa} parton distribution functions (PDFs) with $\alpS(m_{\PZ}) = 0.130$ are used in all generated samples, where \alpS is the strong coupling and $m_{\PZ}$ is the \PZ boson mass.
Backgrounds from top quark-antiquark pair production, \PW{} boson production in association with jets, as well as diboson production, are estimated to contribute less than 1\% to the analysis phase space from MC and are not included.

All generated samples are passed through a detailed simulation of the CMS detector using \GEANTfour~\cite{Agostinelli:2002hh}.
To simulate the effect of pileup, multiple inelastic events are generated using \PYTHIA{8} and superimposed on the primary interaction events.
The MC simulated events are scaled to reproduce the distribution of the number of interactions observed in the experimental data.
The {\mgpy} and \HERWIGpp simulations have additional selection requirements to ensure that the energy scale of the generated primary interaction events is greater than those of the overlapping pileup events.
This can affect the MC \pt distribution slope at low \pt, within the total MC uncertainty; however, this has a negligible effect on the unfolded results and uncertainties.

Additional samples of \PZ boson production in association with jets and QCD multijet production, without detector simulation, are compared with the unfolded experimental data distributions.
Predictions are computed at LO with \PYTHIA v8.243 with the CP2 and CP5 tunes~\cite{Sirunyan:2019dfx}, \HERWIG v7.1.4~\cite{Bellm:2015jjp} at LO with angular-ordered showering and the CH3 tune~\cite{Sirunyan:2020pqv}, and with \SHERPA v2.2.10~\cite{Bothmann:2019yzt, Krauss:2001iv,Gleisberg:2008fv} at LO, and also with two (one) extra jets for QCD multijet (\Zjet) production with the CKKW matching procedure~\cite{Catani:2001cc,Krauss:2002up,Schalicke:2005nv}.
The parton shower in \SHERPA is based on the Catani--Seymour dipole factorization~\cite{Schumann:2007mg}, and hadrons are formed by a modified cluster hadronization model~\cite{Winter:2003tt}.

As discussed in Refs.~\cite{Sirunyan:2018asm,Sirunyan:2019dfx,Sirunyan:2020pqv,Reichelt:2017hts}, the choice of $\alpS(m_{\PZ})$ used in the final-state shower evolution in MC event generators has a significant impact on the shape of jet substructure observables for quark and gluon jets. Whereas the {\mgpy}, \HERWIGpp, and \PYTHIA{8} CP2 predictions use larger values of $\alpS(m_{\PZ})$ for the final-state showering of 0.1383, 0.127, and 0.130, respectively, the \PYTHIA{8} CP5, \HERWIG CH3, and \SHERPA predictions all use a common value of 0.118.
The \PYTHIA{8} CP2 and \PYTHIA{8} CP5 tunes differ mainly in their choice of $\alpS(m_{\PZ})$ that is also used in the underlying event model.
They were obtained by tuning the other parameters controlling the underlying event in \PYTHIA{8} (except $\alpS(m_{\PZ})$) to describe LHC data~\cite{Sirunyan:2019dfx}.
Their comparison thus reflects the difference due to the choice of $\alpS(m_{\PZ})$, while keeping a good description of the underlying event.
In addition to the choice of $\alpS(m_{\PZ})$, the shapes of jet substructure observables are sensitive to the description of gluon splitting to quarks and colour reconnection in the final state~\cite{Gras:2017jty}.

\section{Event selection}
\label{sec:selection}

The measurement of the generalized angularities (Fig.~\ref{fig:lambdadiagram}) is carried out using two event samples: a \Zjet sample that is enriched in quark jets, and a dijet sample that is enriched in gluon jets.
To calculate each angularity for AK4 and AK8 jets, the following event selection is carried out with the AK4 and AK8 jet collections, respectively.
The \Zjet selection requires at least two reconstructed muons with $\pt>26\GeV$, which results in a trigger efficiency exceeding 99\%.
A \PZ boson candidate is reconstructed from two oppositely-charged muons with $\abs{\eta}<2.4$ and a dimuon invariant mass within 20\GeV of $m_{\PZ}$, requiring $\pt^{\PZ}>30\GeV$.
At least one reconstructed jet with $\pt>30\GeV$ and $\abs{y}<1.7$ that does not overlap with the muons of the \PZ boson candidate ($\Delta R(\jet_1, \PGm_{1,2}) > R = 0.4$ or 0.8 for AK4 or AK8 jets, respectively) is required.
The requirement on the jet rapidity ensures that jets are fully within the tracker coverage.
The \PZ boson candidate with transverse momentum $\pt^{\PZ}$, and the jet with the largest \pt, with transverse momentum $\pt^{\jet_1}$, are required to have a \pt asymmetry $\abs{\pt^{\jet_1} - \pt^{\PZ}} / (\pt^{\jet_1} + \pt^{\PZ}) < 0.3$, and an azimuthal separation $\Delta \phi(\jet_1, \PZ) > 2$ radians.
The requirements on \pt asymmetry and azimuthal separation are introduced to restrict the measurement to a phase space dominated by two-body scattering where a single parton is produced together with the \PZ boson in the hard scattering process.
In this phase space large contributions from higher order QCD corrections, arising when $\pt^{\jet_1}$ is significantly larger than $\pt^{\PZ}$~\cite{Rubin:2010xp}, are suppressed.
The substructure of this jet is measured.

The dijet selection requires at least two reconstructed jets with $\pt>30\GeV$ and $\abs{y}<1.7$.
The two jets with largest transverse momentum, $\pt^{\jet_{1,2}}$, are required to have a transverse momentum asymmetry $\abs{\pt^{\jet_1} - \pt^{\jet_2}} / (\pt^{\jet_1} + \pt^{\jet_2}) < 0.3$, and an azimuthal separation larger than 2 radians.
The angularities are calculated for these two jets.
In the following, the jet with smaller $\abs{y}$ is labelled as the central jet, whereas the jet with larger $\abs{y}$ is labelled as the forward jet.

Table~\ref{tab:selections} summarizes the selection criteria of the \Zjet and dijet event samples.
The detector-level distributions are unfolded to particle-level, \ie only considering final-state particles.
At particle level the same selection criteria are applied except that the muons of the \PZ boson candidate are removed from the particles used as input to the jet clustering, instead of requiring $\Delta R(\jet_1, \PGm_{1,2}) > R$.
Muons at particle level are not ``dressed'', \ie radiated photons are not considered part of the muon candidate, since this has a negligible effect on the muons.
Although the detector resolution of jet $\abs{y}$ and $\phi$, as well as that of the muon \pt, is small enough to maintain equivalent particle- and detector-level selection criteria, the resolution of jet \pt is significantly worse.
The measurement is defined for jets with $\pt>50\GeV$ at particle level. However, because of resolution effects, jets with $\pt \sim 50\GeV$ at particle level can migrate to lower transverse momentum bins at detector level. The unfolding procedure is therefore carried out for jets with $\pt>30\GeV$ to minimise modelling uncertainties associated with such migrations.

\begin{table}[htb]
\centering
\topcaption{Summary of the selection criteria for the \Zjet and dijet event samples.}
\begin{tabular}{ll}
Sample & Event selection \\
\hline
\Zjet & $\geq$2 muons with $\abs{\eta}<2.4$ and $\pt^{\mu}>26\GeV$ \\
& Opposite-charge muons \\
& $\abs{m_{\mu\mu} - m_{\PZ}} < 20\GeV$ \\
& $\pt^{\PZ{}} > 30\GeV$ \\
& $\geq$1 jet with $\abs{y}<1.7$ and $\pt^{\jet}>30\GeV$, \\
& $\Delta R(\jet_1, \mu_{1,2}) > R =$ 0.4 or 0.8 \\
& $\Delta \phi(\jet_1, \PZ) > 2$ \\
& $\abs{\pt^{\jet_1} - \pt^{\PZ{}}} / (\pt^{\jet_1} + \pt^{\PZ{}}) < 0.3$ \\
Dijet & $\geq$2 jets with $\abs{y}<1.7$ and $\pt^{\jet}>30\GeV$ \\
(central and & $\Delta \phi (\jet_1, \jet_2) > 2$ \\
forward) & $\abs{\pt^{\jet_1} - \pt^{\jet_2}} / (\pt^{\jet_1} + \pt^{\jet_2}) < 0.3$ \\
\end{tabular}
\label{tab:selections}
\end{table}

\begin{figure}[!htp]
    \centering
    \includegraphics[width=0.49\textwidth]{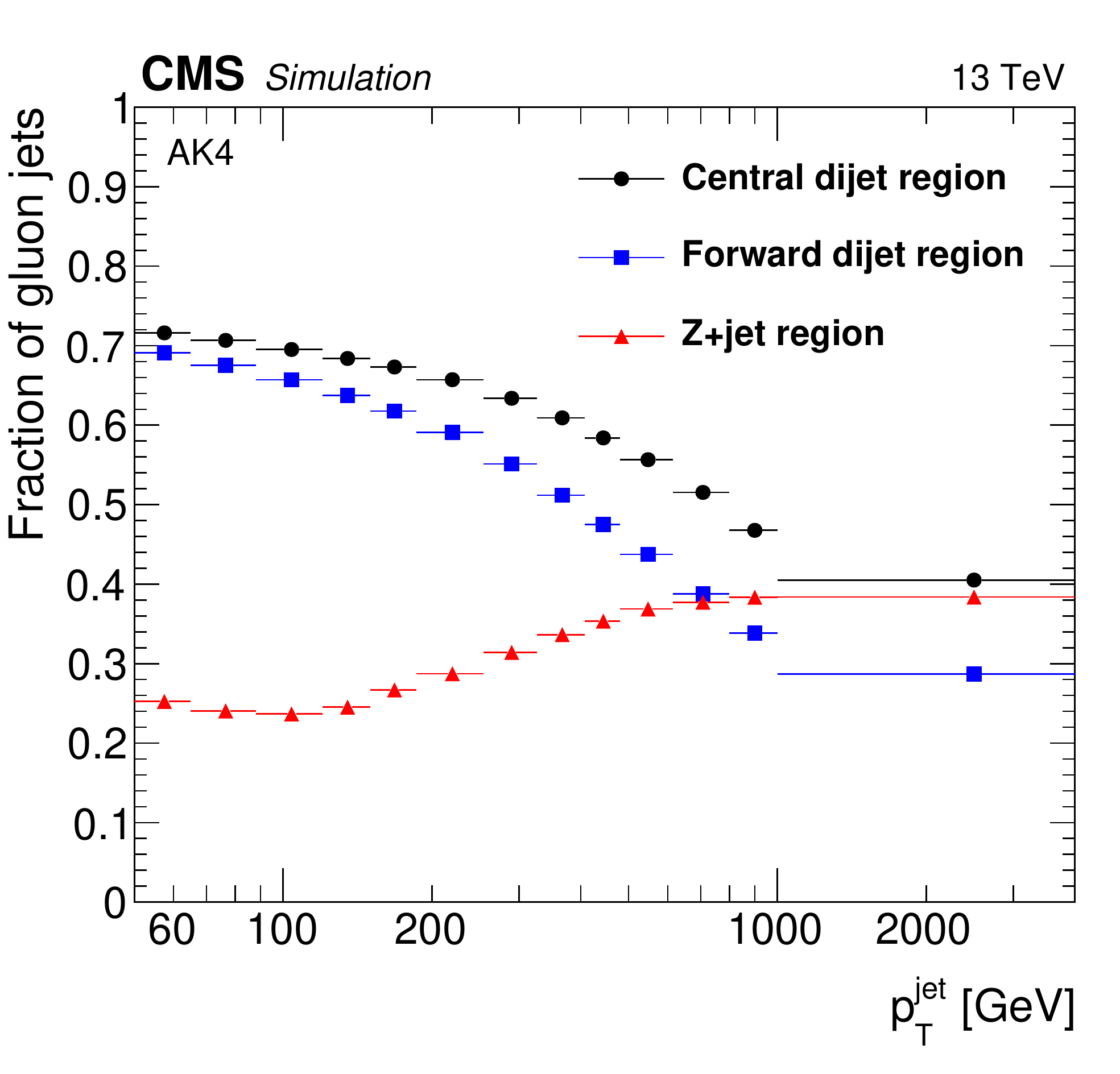}
    \caption{Fraction of AK4 jets that are gluon jets in the \Zjet sample (red triangles), and the central (black circles) and forward (blue squares) dijet samples.
    The statistical uncertainty is negligible in the simulation.}
    \label{fig:flavFractions}
\end{figure}

We quantify the fraction of quark and gluon jets in these event samples using the {\mgpy} simulation.
As discussed in Ref.~\cite{Gras:2017jty}, there is no unique way of defining a quark or gluon jet, and our choice of approach is as follows.
The parton from which a jet originates is obtained by first clustering jets from all generated partons (from the matrix element and the shower up to hadronization) together with the generated final-state particles, where the four-momenta of the generated partons are scaled by a very small number.
The jets clustered in this way are almost identical to those clustered without generator partons because the added partons, with extremely small energy, do not affect the clustered jet momentum.
The quark or gluon with highest \pt in the jet then defines the origin of the jet.
Figure~\ref{fig:flavFractions} shows the resulting fraction of AK4 jets which are gluon jets in the \Zjet sample, and central and forward dijet samples as a function of jet \pt.
The \Zjet sample is enriched in quark jets, making up 64--76\% of the jets in the sample.
The central and forward dijet samples are dominated by 69--72\% gluon jets at lowest \pt, and by 55--68\% quark jets at highest \pt.
The central dijet jets have a consistently higher fraction of gluon jets than the forward dijet jets across the entire \pt range.
The fraction of gluon jets was also computed with \HERWIGpp, \PYTHIA{8} CP5, \PYTHIA{8} CP2, \HERWIG{}7 CH3 and \SHERPA.
Although the gluon jet fraction in all generators agrees with {\mgpy} within 6\% in the dijet samples, up to 25\% lower gluon fractions are predicted in the \Zjet sample.
The spread reflects the rather large uncertainties associated with LO MC generator predictions.

In Fig.~\ref{fig:flavFractions}, we consider all quark flavours, except for top, to identify quark jets.
Typically, however, quark jets are considered to be jets initiated only by light-flavour quarks ($\PQu, \PQd, \PQs$).
The \Zjet and dijet samples also have a small contribution of 5--15\% from jets initiated by heavy-flavour quarks ($\PQb, \PQc$), whose behaviour was measured in Ref.~\cite{Sirunyan:2018asm}, and is in between that of light-flavour quark jets and gluon jets.
Their presence will therefore reduce the expected differences between jets from the \Zjet and dijet samples.

\begin{figure}[!htp]
    \centering
    \includegraphics[width=0.49\textwidth]{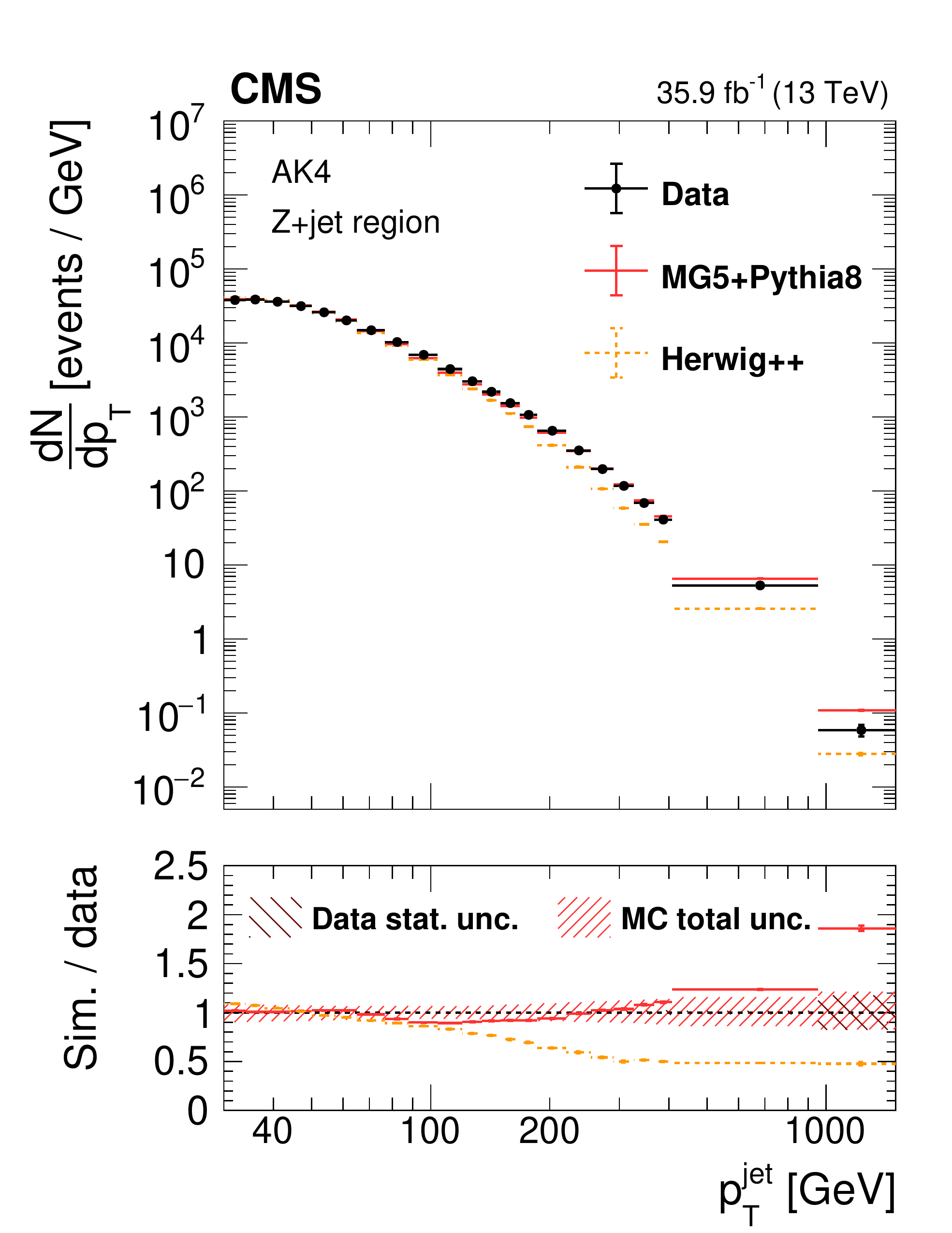}
    \includegraphics[width=0.49\textwidth]{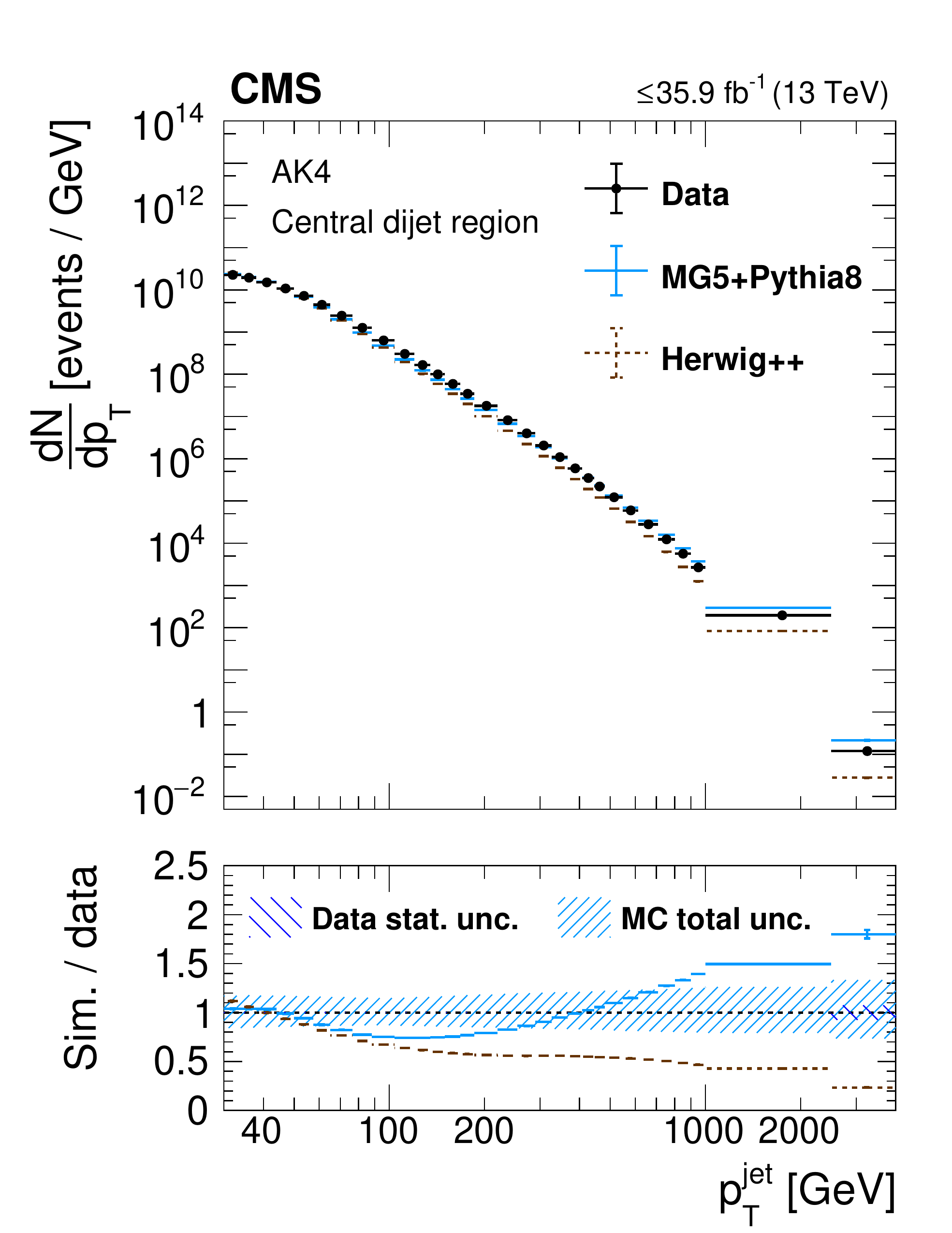}
    \caption{Data to simulation comparisons of the jet \pt in the \Zjet (left) and central dijet (right) regions.
     The error bars correspond to the statistical uncertainties in the experimental data.
     The coarse-grained hashed region in each ratio plot indicates the statistical uncertainty in the experimental data, and the fine-grained hashed region represents the total uncertainty in the MC prediction.
    }
    \label{fig:ptDataMcSpectrums}
\end{figure}

The \pt distribution in the CMS data and simulation for the \Zjet and central dijet AK4 jets is shown in Fig.~\ref{fig:ptDataMcSpectrums}.
The data distribution in bins of jet \pt collected with prescaled triggers (see Table~\ref{tab:triggers}) is multiplied by the corresponding prescale factors to reproduce the smoothly falling physical spectrum.
The data distribution is compared with predictions from two generator configurations, where the total number of events in the simulations has been scaled to match the data distribution.
In both selection regions, the {\mgpy} simulation provides a better description of the experimental data, and is therefore used to estimate the detector response consistently in both samples in this paper, whereas \HERWIGpp is used to evaluate the systematic uncertainties.
The disagreements observed in the \Zjet and central dijet \pt distributions indicate a sensitivity to higher order corrections not fully modelled in these two generators.
These differences have, however, a smaller effect on the jet substructure distributions that are measured in bins of jet \pt.

\section{Jet substructure observables}
\label{sec:observables}

For the AK4 and AK8 jets of interest, we compute and measure five generalized angularities as defined in Equation~\ref{angularities} that discriminate quark and gluon jets.
Of these variables, only LHA, width, and thrust are IRC-safe as a result of having $\kappa = 1$~\cite{Larkoski:2014uqa}.
However, multiplicity and \pTD have been widely used for quark and gluon discrimination~\cite{CMS-PAS-JME-13-002,CMS-PAS-JME-16-003}.
For $\beta > 1$, the anti-\kt jet axis is used to calculate $\DR$.
For $\beta \le 1$, $\DR$ is calculated using the jet axis constructed by the winner-takes-all (WTA) method~\cite{Larkoski:2014uqa}.
The use of the WTA axis significantly simplifies theoretical calculations of the angularities, whereas for angularities with $\beta > 1$ computation is feasible with both axis definitions.
However, using the anti-\kt jet axis results in an observable that is more akin to the jet mass~\cite{Sirunyan:2018xdh}.
To calculate the WTA axis, the constituents of the anti-\kt jet are reclustered using the \pt-based WTA scheme.
The resultant jet has the $(y, \phi)$ of the constituent with the largest \pt.
The calculation of multiplicity includes only constituents with $\pt>1\GeV$, at both the detector and particle levels.

Each of the angularities is designed to emphasize a particular feature of the jet.
The LHA, width, and thrust include both the momentum fraction and the angular distribution of the constituents within the jet.
Since the weighting of the angular distribution differs across these variables, comparing them can highlight differences in the constituent positions within the jet.
We expect gluon jets to have, on average, larger values of these angularities due to the larger number of constituents that are further from the jet axis.
In contrast, \pTD places more value on those higher-momentum constituents, irrespective of their position in the jet.
With gluon jets typically having more lower-momentum constituents, we therefore expect them to generally have smaller values of \pTD.
Whereas the three IRC-safe angularities are particularly sensitive to the modelling of perturbative emissions in jets, the other two have larger contributions from nonperturbative effects and are thus subject to larger uncertainties in their predictions and measurements.

In this paper, we measure each substructure observable with various configurations.
Each quantity is computed in multiple bins of jet \pt, and for two different jet radius parameters, $R = 0.4$ and 0.8.
For each quantity, we define a variant where the observables is calculated using only the charged constituents in anti-\kt algorithm (``charged'').
While observables computed with both charged and neutral constituents can be described more easily from first-principle calculations, the charged variants can be measured with a better resolution as a result of the high efficiency and precision of the tracking detector.

Additionally, we compute a groomed variant of each observable, where the jet is reclustered with the Cambridge--Aachen (CA) algorithm~\cite{Dokshitzer:1997in}, and then groomed using the modified mass-drop algorithm~\cite{Dasgupta:2013ihk,Butterworth:2008iy}, known as the \emph{soft drop} algorithm~\cite{Larkoski:2014wba}.
This splits the jet into two subjets by undoing the last step of the CA jet clustering.
It regards the jet as the final soft drop jet if the two subjets satisfy the condition:
\begin{linenomath}
\begin{equation}
 \frac{ \min ( \pt^{(1)}, \pt^{(2)} ) }{\pt^{(1)} + \pt^{(2)}  } > z_{\text{cut}} \Big(  \frac{ \Delta R_{12} }{R} \Big) ^ {\beta_{\text{sd}}},
\end{equation}
\end{linenomath}
where $\pt^{(1)}$ and $\pt^{(2)}$ are the transverse momenta of the two subjets, $R$ is the jet radius parameter, $\Delta R_{12} = \sqrt{(\Delta y_{12})^2 + (\Delta \phi_{12})^2}$ is the distance between the two subjets, and $z_{\text{cut}} $ and $\beta_{\text{sd}}$ are tunable parameters of the soft drop algorithm, set to $z_{\text{cut}} = 0.1$ and $\beta_{\text{sd}} = 0$ in this study.
If the condition is not met, the subjet with the lower \pt is rejected.
The declustering procedure is repeated splitting the higher \pt subjet into two subjets by undoing another CA clustering step, iterating until the condition is met.
This grooming procedure removes soft and wide-angle radiation from the jet, thereby making the jet substructure observables more resilient to effects from pileup, underlying event, and initial-state radiation.
In all cases, the jet \pt used for binning is taken from the original ungroomed anti-\kt jet (\ie using charged+neutral constituents, before calculating the WTA axis, and before grooming).
A summary of all variations of the observables and the \pt ranges measured in this paper is given in Table~\ref{tab:observables}.
Whereas the measurements are carried out for all variations of the observables and made public in HEPData record~\cite{hepdata}, this paper will focus on representative distributions, typically taking the ungroomed charged+neutral distribution with $R$=0.4 at $120<\pt<150\GeV$ as a reference for comparisons.
This \pt range is chosen as a compromise between lowest \pt with highest enrichment in gluons for the dijet central sample and higher \pt yielding lower statistical uncertainties for the dijet sample (see Table~\ref{tab:triggers}) and lower systematic uncertainties because of better jet energy resolutions.

\begin{table}[htb]
\centering
\topcaption{Summary of the variants of observables measured in this paper.}
\begin{tabular}{lc}
Dimension & Variants \\
\hline
Region & \Zjet vs. central dijet vs. forward dijet \\
Observable $\lambdakb$ & LHA, width, thrust, multiplicity, \pTD \\
Jet \pt & $50<\pt<65\GeV$, \ldots, $\pt>408\GeV$ for \Zjet \\
 & $50<\pt<65\GeV$, \ldots, $\pt>1000\GeV$ for dijet \\
Jet radius parameter $R$ & 0.4 vs. 0.8 \\
Constituents & Charged+neutral vs. charged \\
Grooming & Ungroomed vs. groomed \\
\end{tabular}
\label{tab:observables}
\end{table}

\section{Unfolding and systematic uncertainties}
\label{sec:unfolding}

The experimental data distributions are unfolded to particle level using unregularized unfolding as implemented in the \textsc{TUnfold} package~\cite{Schmitt:2012kp}.
The particle-level distribution and covariance matrix are obtained by minimizing
$\chi^{2}_A = \left(y-Ax\right)^{T} V_{yy}^{-1} \left(y-Ax\right)$,
where $A$ is the particle-to-reconstructed phase space migration matrix, $V_{yy}$ is an estimate of the variance of the observations $y$, and $x$ is an estimator of the true particle-level values.
Here, $V_{yy}$ is a diagonal matrix formed by the square of the bin errors.
The migration matrix in the 2D phase space of $(\pt, \lambdakb)$ is derived from the {\mgpy} simulation, matching particle-level jets with detector-level jets by requiring an angular separation of less than half of the jet radius parameter.
Before the detector-level distribution is unfolded, background from misreconstructed jets, \ie jets that pass the reconstruction criteria, but fail the particle-level selection, are subtracted.
The proportion of misreconstructed jets ranges from ${<}10\%$ for jets at 50\GeV to ${<}1\%$ at 1\TeV.
The contribution from other standard model processes was negligible.
The migration matrix also includes those particle-level jets that fail the reconstruction criteria, to correct for the reconstruction inefficiency.

The binning of the migration matrix includes the detector resolution.
We define purity as the fraction of reconstructed events that are generated in the
same $(\pt, \lambdakb)$ bin, and stability as the fraction of generated events that are reconstructed in the same
bin.
Both quantities are $\geq$50\% in most bins.
The detector-level bins have half the width of the particle-level bins in both \pt and \lambdakb, to guarantee an over constrained system for the minimization carried out during unfolding.

Various sources of systematic uncertainty are considered, which are treated as uncorrelated among each other, but correlated in the $(\pt, \lambdakb)$ plane.
These are divided into experimental sources, uncertainties related to the theoretical model used to derive the migration matrix, and inherent unfolding uncertainties.
Experimental and physics model uncertainties are treated by constructing variations in the migration matrix, and propagating the changes to the final unfolded 1D distribution.

Uncertainties in the calibration of the jet energy scale (JES) and jet energy resolution (JER) measurement~\cite{Khachatryan:2016kdb} are included by varying the jet \pt when constructing the migration matrix.
Furthermore, we account for variations in the measurement of the particle-flow candidates, which are propagated to the computation of the jet substructure observables in the migration matrix.
These include variations of the charged-hadron momentum scale by $\pm$1\%, photon energy scale by $\pm$1\%, and neutral-hadron energy scale by $\pm$3\%~\cite{CMS-PRF-14-001, Khachatryan:2016kdb} that are estimated from single-particle response measurements~\cite{CMS-PAS-JME-10-008}.
We find uncertainties related to the angular resolution of the particle-flow reconstruction and charged-hadron reconstruction efficiency and momentum resolution are negligible.
The simulation of pileup is tuned to match the particle multiplicities and \pt spectra observed in data with a dedicated MC tune~\cite{Sirunyan:2019dfx}.
We estimate the uncertainty in jet \pt and substructure observables related to pileup by reweighting the simulated events within the uncertainty of the distribution of the average number of interactions per bunch crossing.
For the \Zjet region, this corresponds to a $\pm$4.6\% shift in the total inelastic cross section~\cite{Sirunyan:2018nqx}, whereas for the dijet region it is a $+4.6$/$-13.8\%$ shift to account for larger variations coming from the use of prescaled triggers.
Since the luminosity collected by prescaled triggers varies over the data-taking periods, the pileup profile differs from the pileup distribution of the full dataset.
Uncertainties related to the inelastic $\Pp\Pp$ model~\cite{Khachatryan:2015pea} used to simulate pileup interactions are assumed to be negligible, because the variations in the model would produce a smaller impact on the charged particle multiplicity than the changes in the average number of interactions per bunch crossing (discussed above).
No uncertainty related to the integrated luminosity is considered since this affects only the overall normalization, and therefore does not affect the normalized distributions presented in this paper.
Uncertainties in the muon reconstruction and identification are negligible.

Uncertainties because of the choices of the factorization and renormalization scales ($\mu_{\mathrm{F}}, \mu_{\mathrm{R}}$) are computed from the envelope of six variations by factors (0.5, 0.5), (0.5,1), (1,0.5), (2,2), (2,1) and (1,2)~\cite{Cacciari:2003fi, Catani:2003zt}.
The uncertainty from the PDFs is determined using the LO NNPDF replica set, where the root-mean-square of 100 pseudo-experiments provided by the PDF set represents the uncertainty.
For the {\mgpy} prediction, these uncertainties related to scale choices and PDFs mainly vary the prediction of jet momenta, whereas the substructure observables are only mildly modified.
The uncertainty in the modelling of the parton shower and hadronization process is estimated using \HERWIGpp to construct an alternative response matrix.
The statistical uncertainty in the \HERWIGpp simulation is small compared to the parton shower and hadronization uncertainty.
This uncertainty is treated as single-sided when building the covariance matrix, since the experimental data distributions are generally enveloped by the {\mgpy} and \HERWIGpp predictions.
We tested that rescaling the \pt spectrum of the \HERWIGpp prediction to match that of the {\mgpy} prediction does not significantly change the estimated uncertainty, confirming that this uncertainty mainly captures differences in the shower and hadronization modelling of substructure rather than the modelling of the \pt spectrum.
We conclude that the mismodeling of the \pt spectrum in Fig.~\ref{fig:ptDataMcSpectrums} has a negligible effect on the result.

The statistical uncertainty in the simulation that is used to derive the migration matrix is propagated through the unfolding and results in a contribution to the covariance matrix.
The statistical uncertainties in the presented distributions correspond to those of the input experimental data, whereas statistical uncertainties in the response matrix are considered as part of the systematic uncertainties.
The unfolding procedure reproduces the particle-level distributions when unfolding a statistically independent sample from the same generator that is used to derive the response matrix.

\begin{figure}[!htb]
    \centering
    \includegraphics[width=0.49\textwidth]{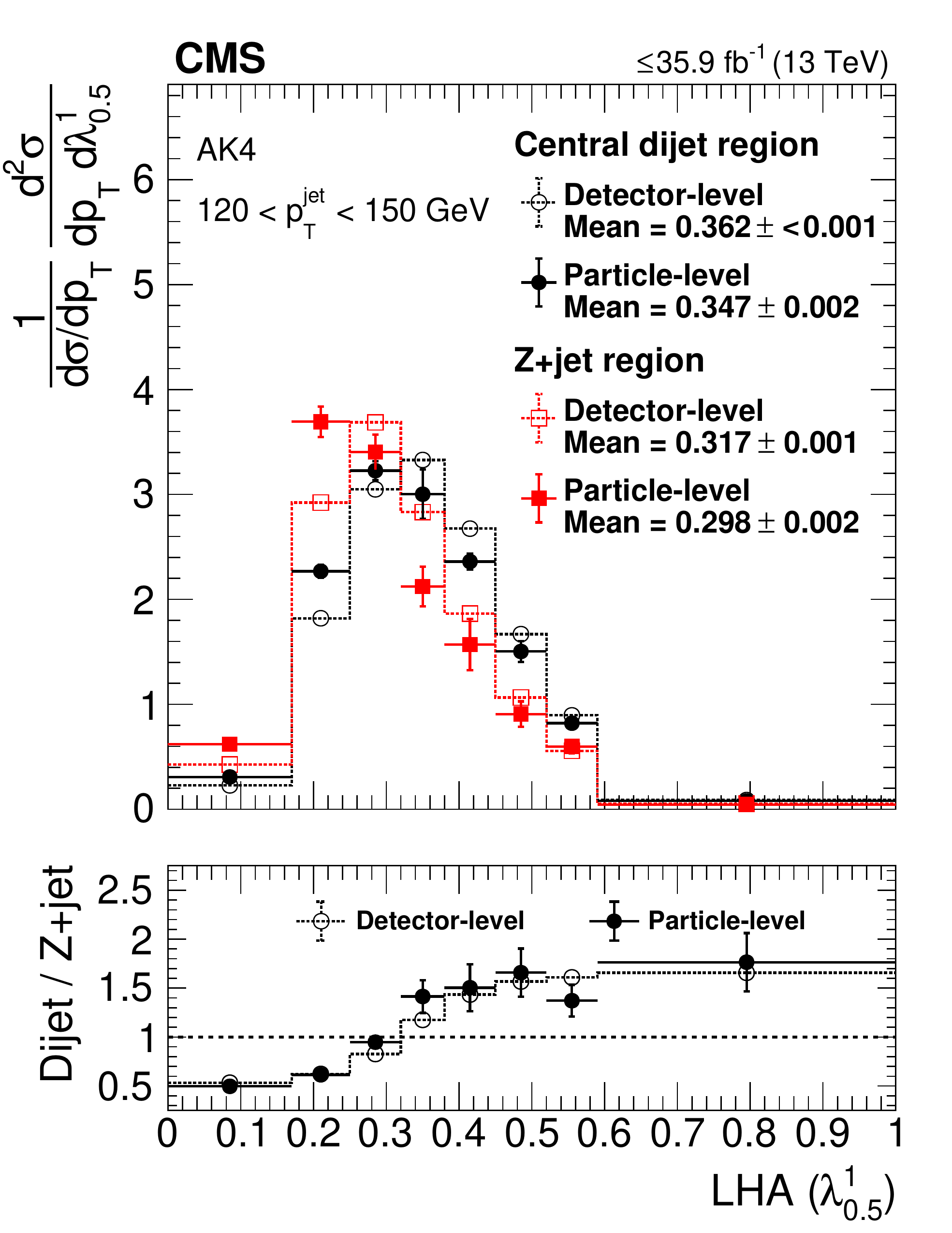}
    \includegraphics[width=0.49\textwidth]{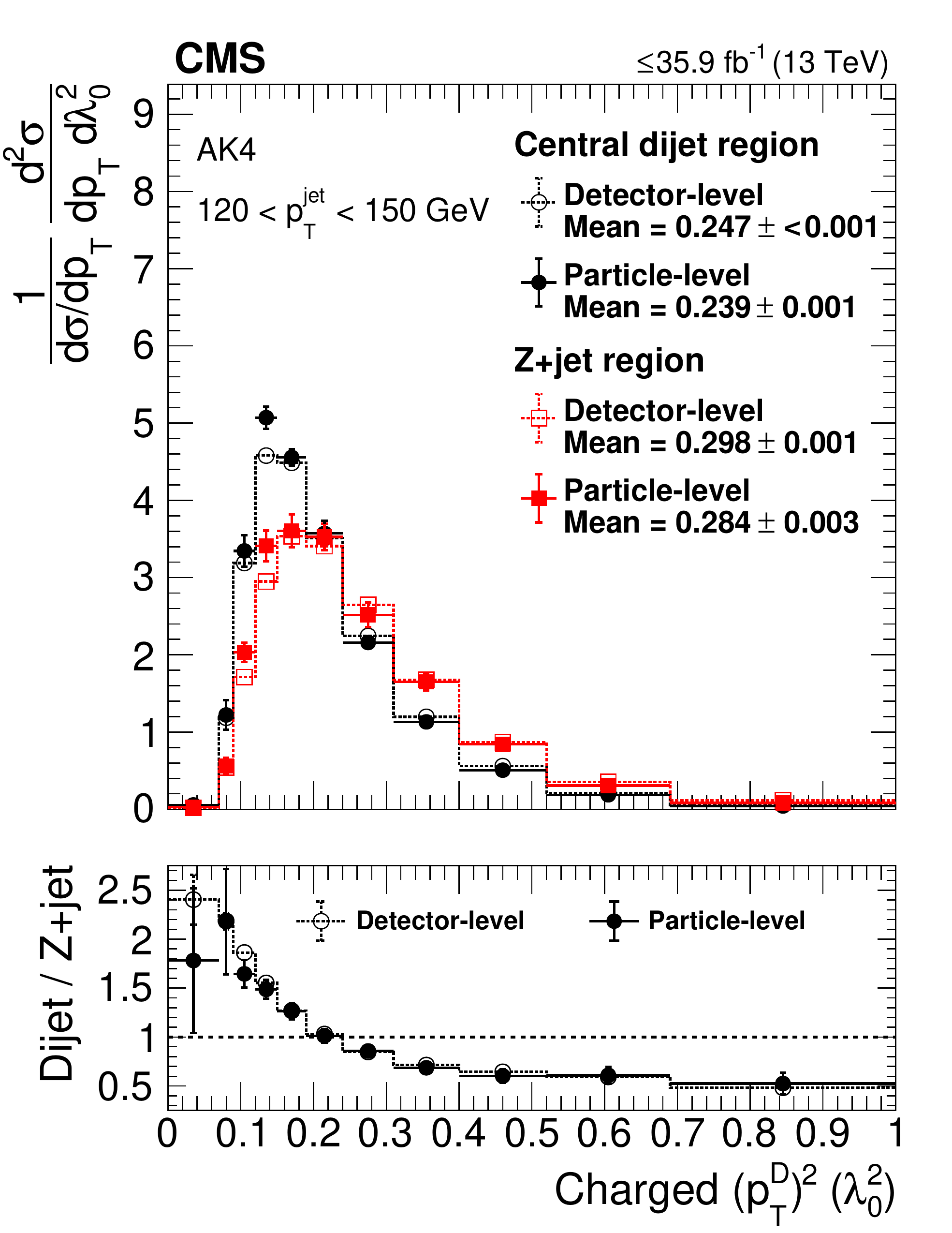}
    \caption{Detector-level and particle-level (unfolded) experimental data distributions of LHA (\lha) (left) using charged+neutral constituents, and \pTD (\ptd) (right) using only charged constituents, for jets with $120<\pt<150\GeV$ in the \Zjet (red) and central dijet (black) regions.
    The detector-level data uncertainties are statistical.
    The particle-level (unfolded) data uncertainties include the systematic components.
    Also shown is the mean of each distribution, calculated from the binned data.
    The ratio plots show the ratio of dijet to \Zjet distributions, for both the detector- and particle-level distributions.}
    \label{fig:substructureDataMcSpectrums}
\end{figure}

Figure~\ref{fig:substructureDataMcSpectrums} shows the distributions of two representative jet substructure observables in experimental data, computed using AK4 jets with $120<\pt<150\GeV$ in the central dijet and \Zjet regions, at both the detector and particle levels.
The discrimination power between quark and gluon jets for each observable can be deduced by comparing the distributions in the \Zjet and central dijet regions that are enriched in quark and gluon jets, respectively.
As previously mentioned, gluon jets exhibit typically larger values of LHA, width, thrust, and multiplicity, and smaller values of \pTD, than quark jets.

The distortion of the distribution by the detector effects can be inferred by comparing the detector-level data distributions with their corresponding particle-level (unfolded) data distributions.
The charged \pTD (Fig.~\ref{fig:substructureDataMcSpectrums}, right) is an example where the difference between detector-level data and particle-level (unfolded) data is small as a result of  considering only charged constituents.
To aid comparison, the mean of each distribution is quoted.
The mean is computed from the binned distribution, which is an approximation to the real mean of the underlying distribution, treating experimental data and simulation consistently, both at detector  and particle levels.
Although the uncertainty quoted in the mean of the detector-level data is purely statistical, the uncertainty quoted for the mean in particle-level (unfolded) data also includes systematic uncertainties.

\begin{figure}[!htb]
    \centering
    \includegraphics[width=0.49\textwidth]{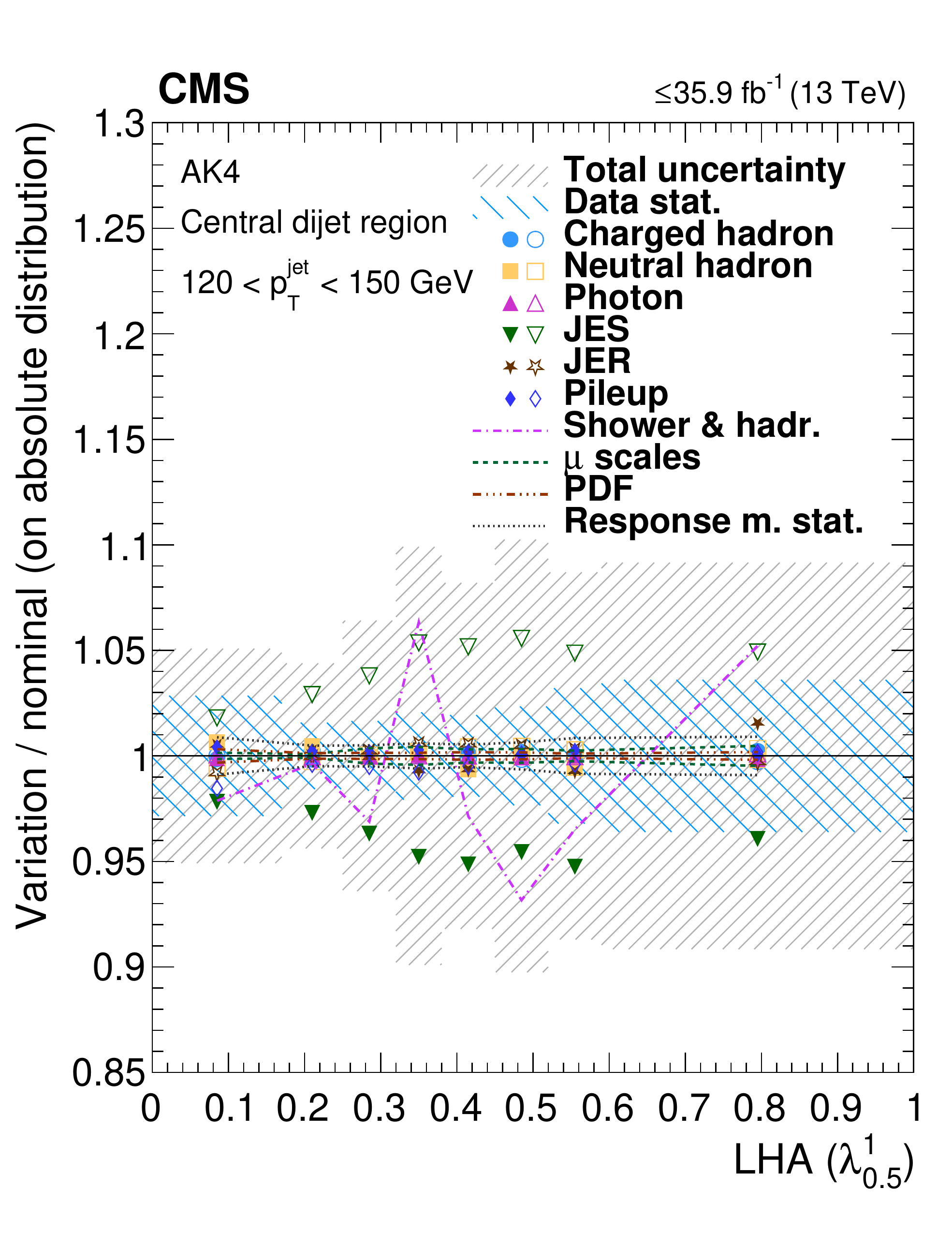}
    \includegraphics[width=0.49\textwidth]{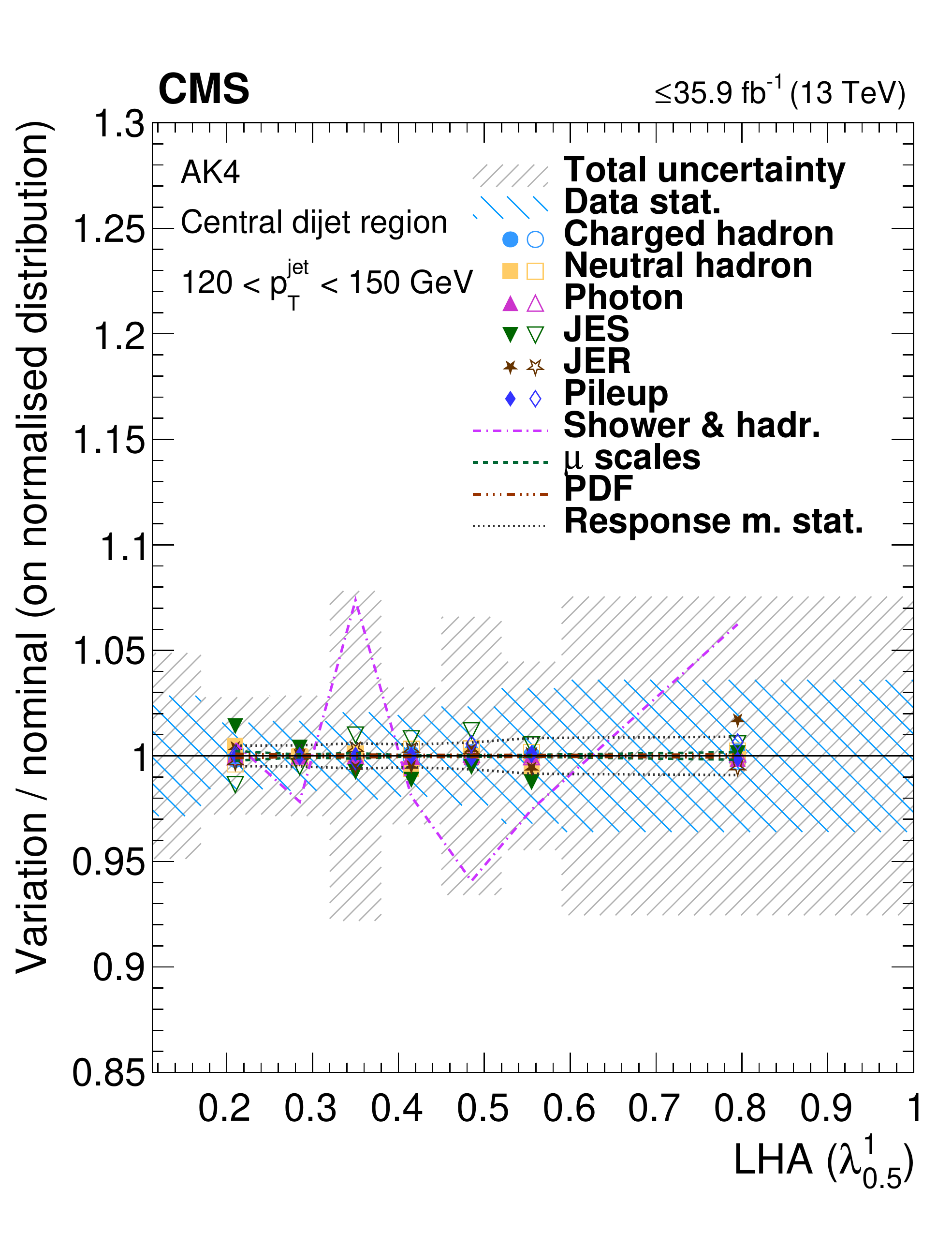}
    \caption{Ratio of the distribution that results from a varied response matrix to the nominal distribution of ungroomed LHA (\lha) for AK4 jets with $120<\pt<150\GeV$ in the central dijet region before (left) and after (right) normalization.
    Filled symbols correspond to upward variations, and the corresponding downward variations are represented by open symbols.
    The coarse-grained hashed region indicates the statistical uncertainty of the experimental data, and the fine-grained hashed region represents the total uncertainty, the sum in quadrature of statistical and systematic uncertainties.
    }
    \label{fig:unfoldedUncertLHA}
\end{figure}

Figure~\ref{fig:unfoldedUncertLHA} shows the size of the considered uncertainties in the ungroomed LHA (\lha) for AK4 jets with $120<\pt<150\GeV$ in the central dijet region before (left) and after (right) the distribution for the given \pt bin is normalized.
Although the JES uncertainty has the largest effect in the yield variations due to migration across the jet \pt boundaries, it plays a limited role for the normalized jet substructure distributions, where the shower and hadronization model uncertainty is typically dominant.
The shape of the shower and hadronization uncertainty is related to some extend to the shape of the LHA (\lha) distribution of \HERWIGpp, exhibiting a lower mean than {\mgpy}.
The total systematic uncertainty is computed from the sum in quadrature of statistical and systematic uncertainties in each bin.

\section{Results and discussion}
\label{sec:results}

\begin{figure}[!htp]
    \centering
    \includegraphics[width=0.49\textwidth]{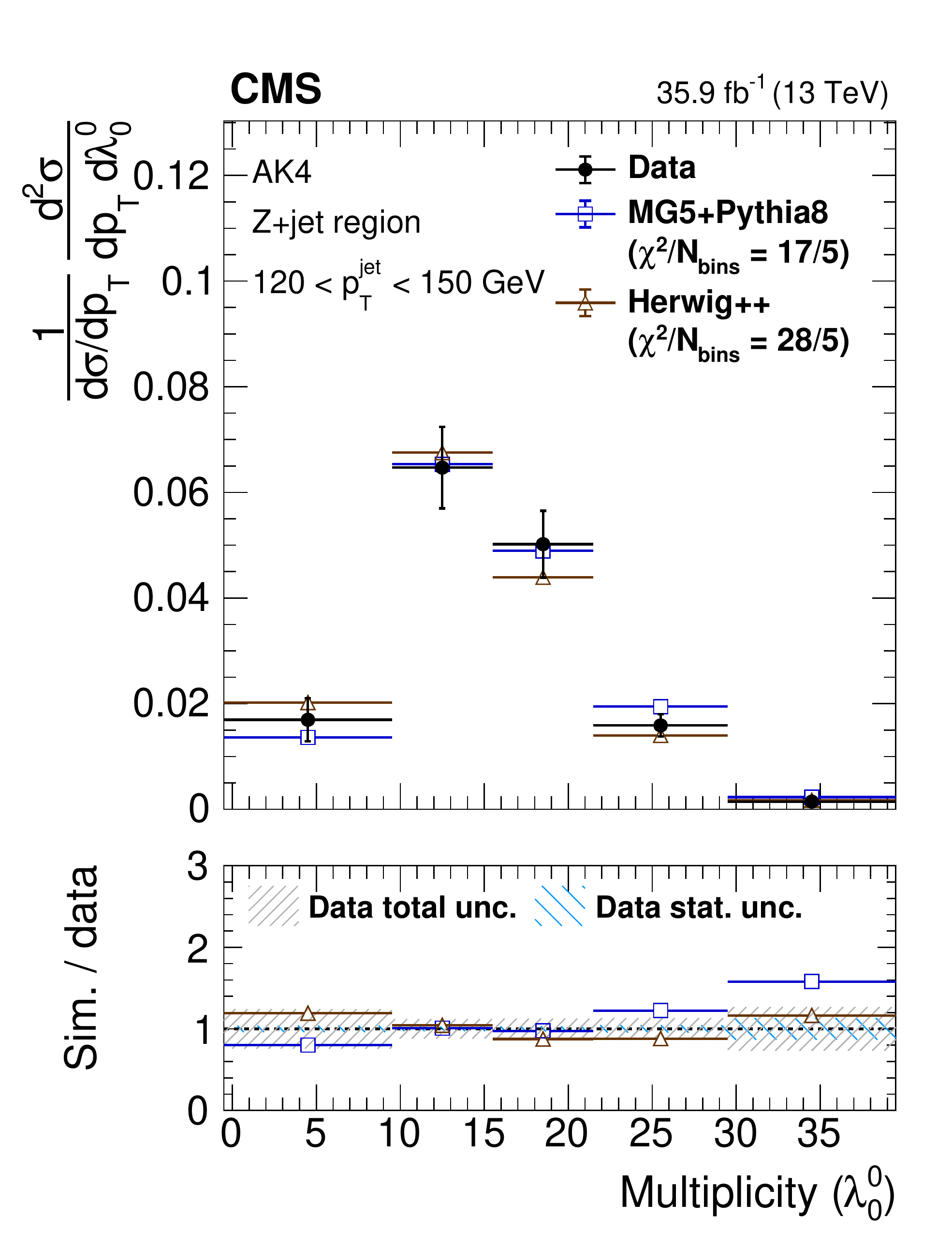}
    \includegraphics[width=0.49\textwidth]{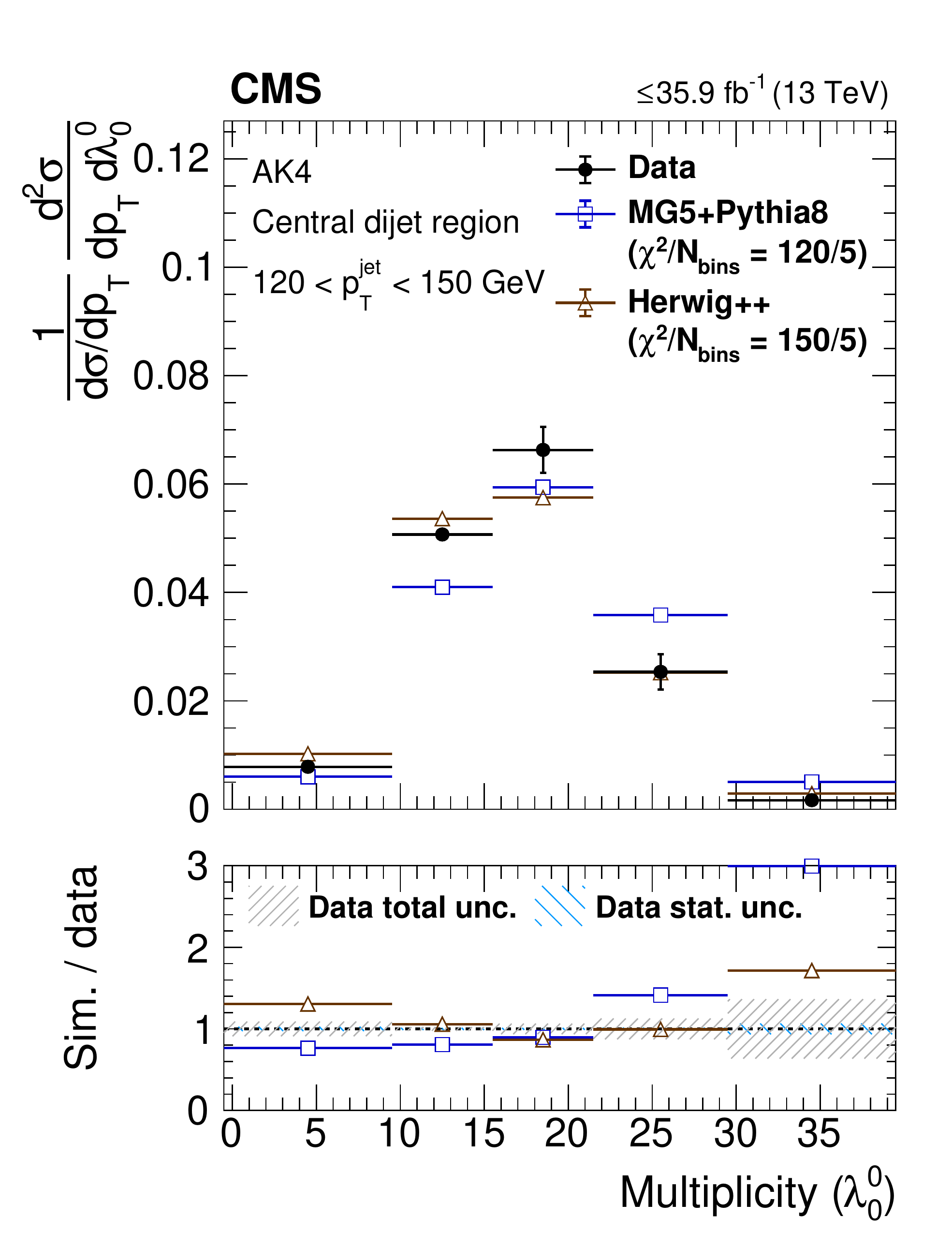} \\
    \includegraphics[width=0.49\textwidth]{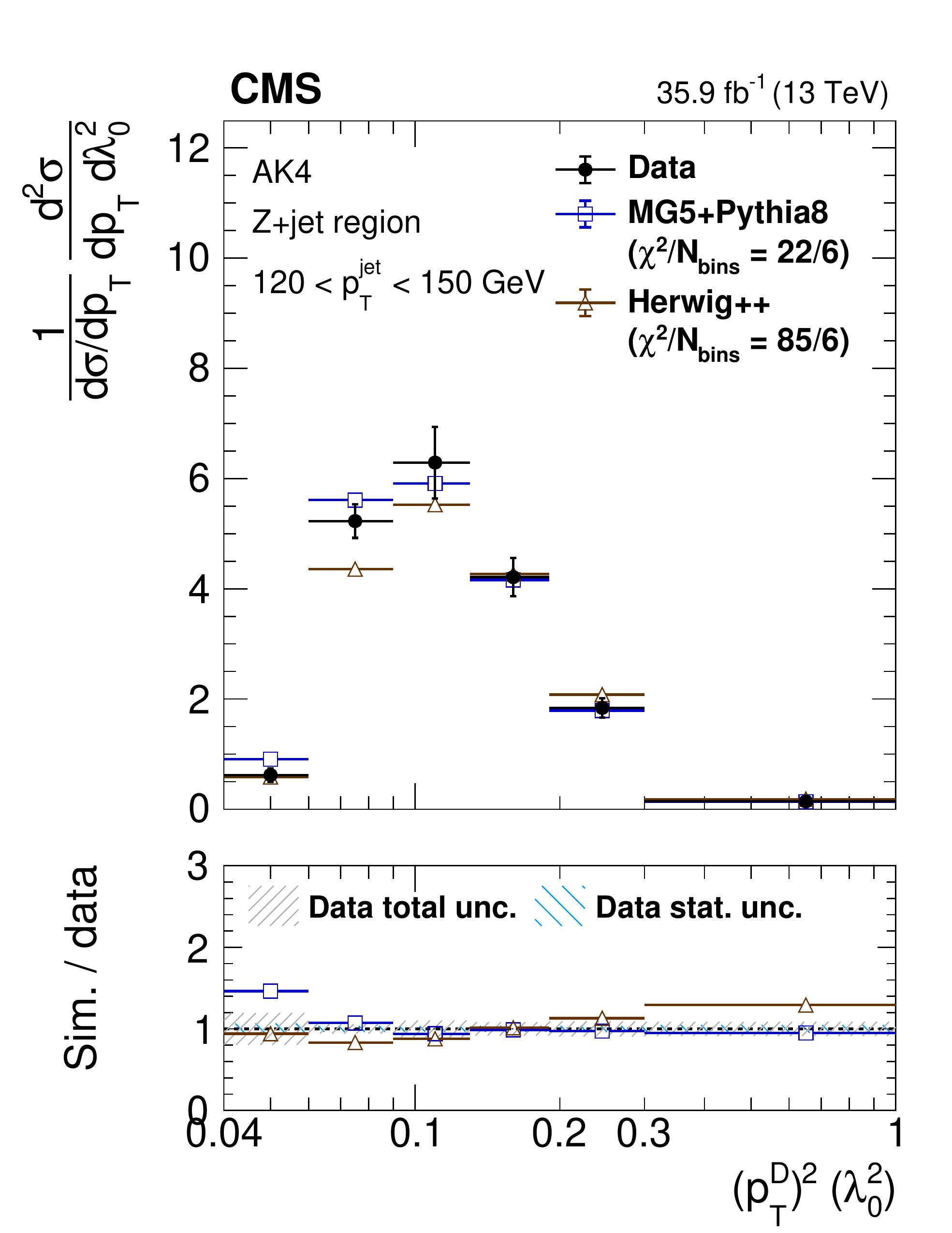}
    \includegraphics[width=0.49\textwidth]{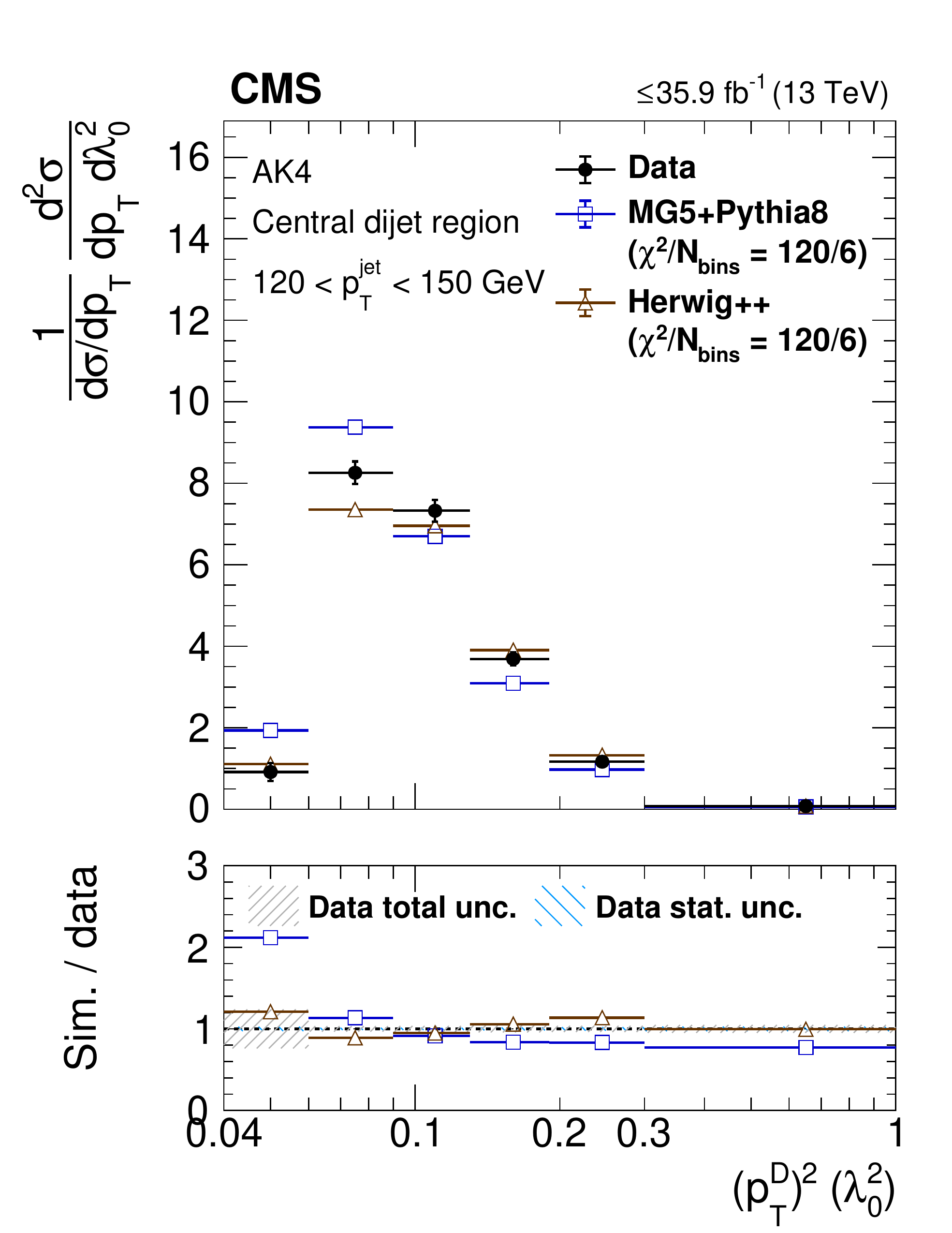}
    \caption{Particle-level distributions of (upper) ungroomed multiplicity (\multi) and (lower) ungroomed \pTD (\ptd) in $120<\pt<150$\GeV in the \Zjet region (left) and central dijet region (right).
    The error bars on the data correspond to the total uncertainties.
    The coarse-grained blue hashed region in the ratio plot indicates the statistical uncertainty of the data, and the fine-grained grey hashed region represents the total uncertainty.
    The lowest bin extends down to $\lambdakb>=0$.
    }
    \label{fig:UnfoldedSpectra4}
\end{figure}

\begin{figure}[!htp]
    \centering
    \includegraphics[width=0.49\textwidth]{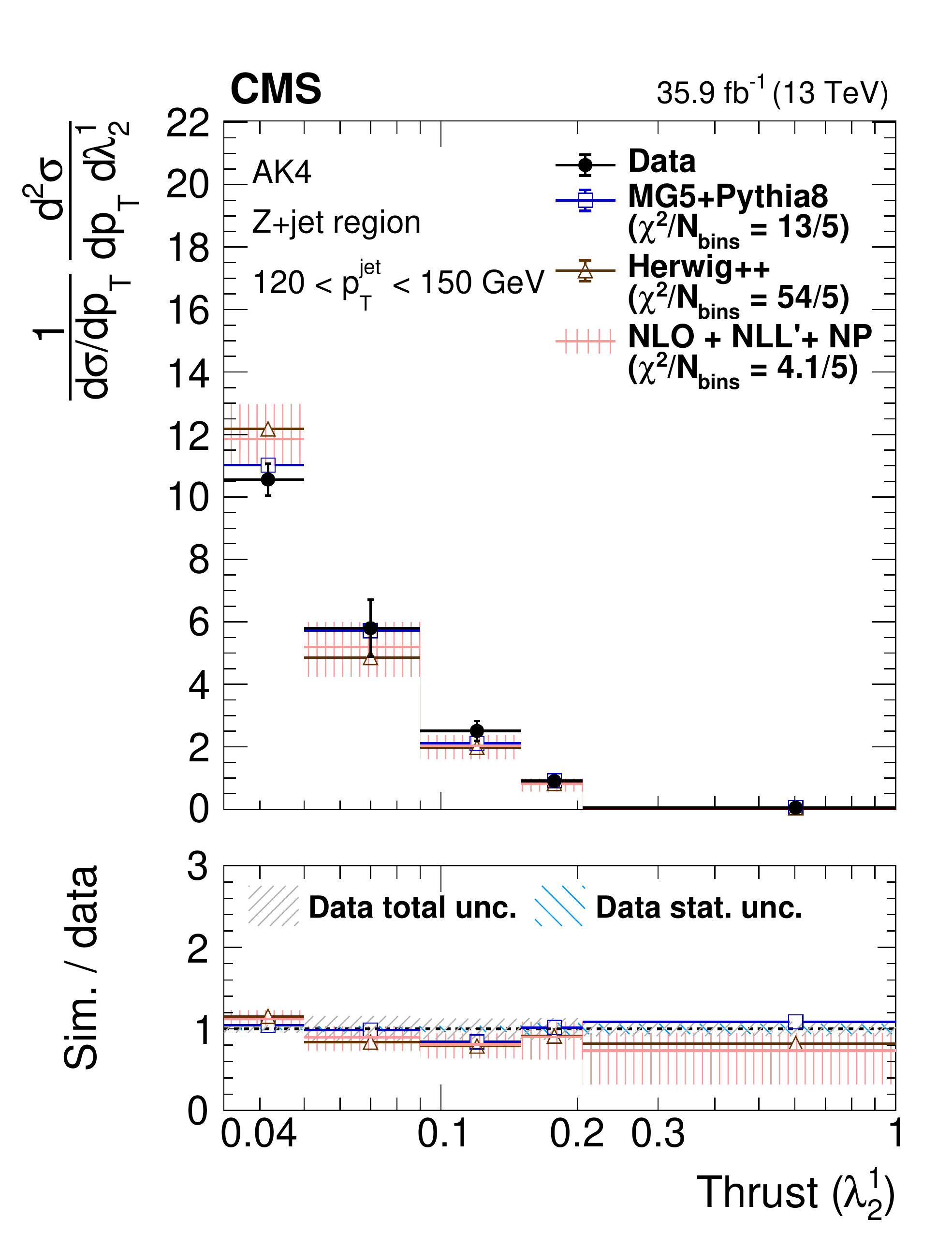}
    \includegraphics[width=0.49\textwidth]{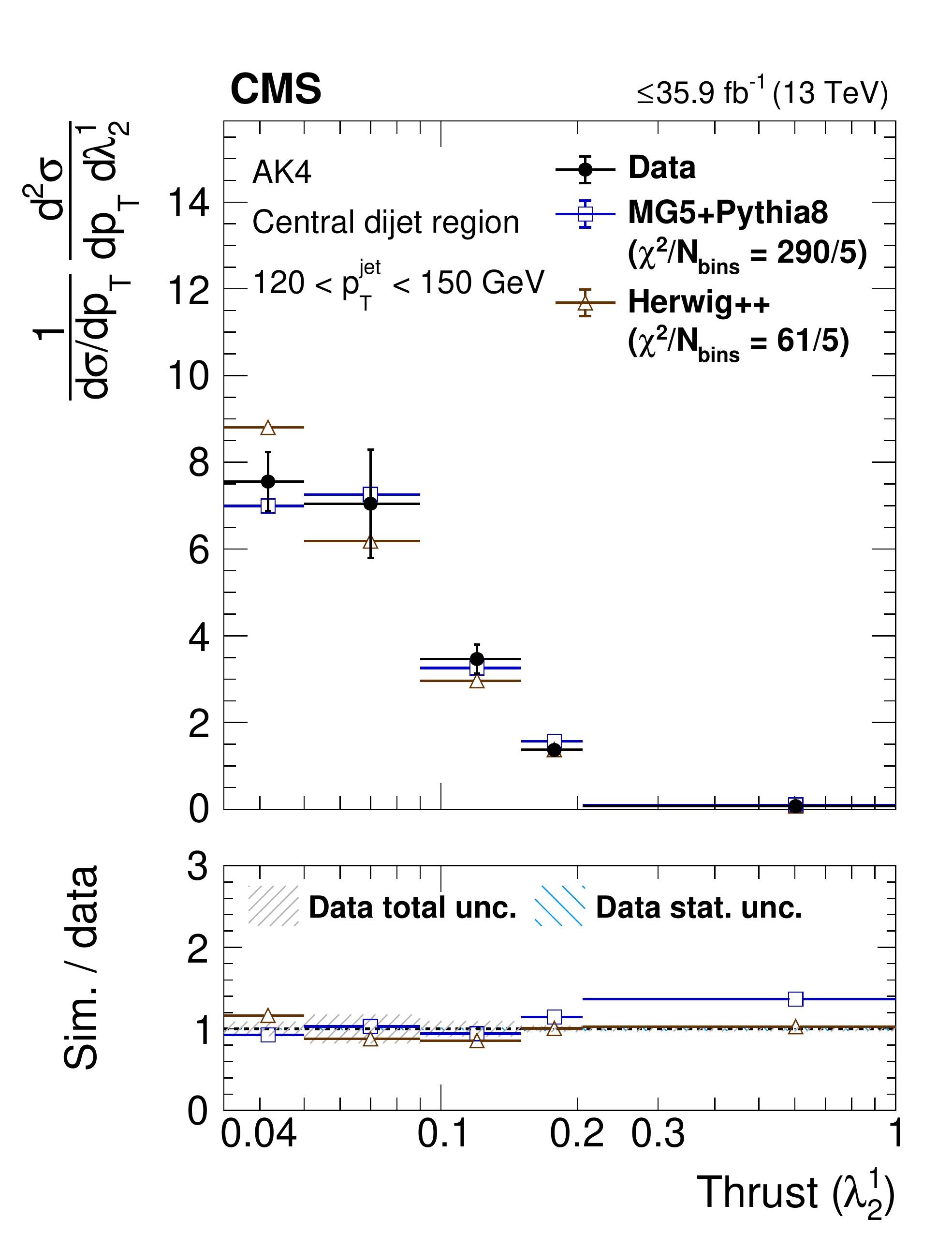} \\
    \includegraphics[width=0.49\textwidth]{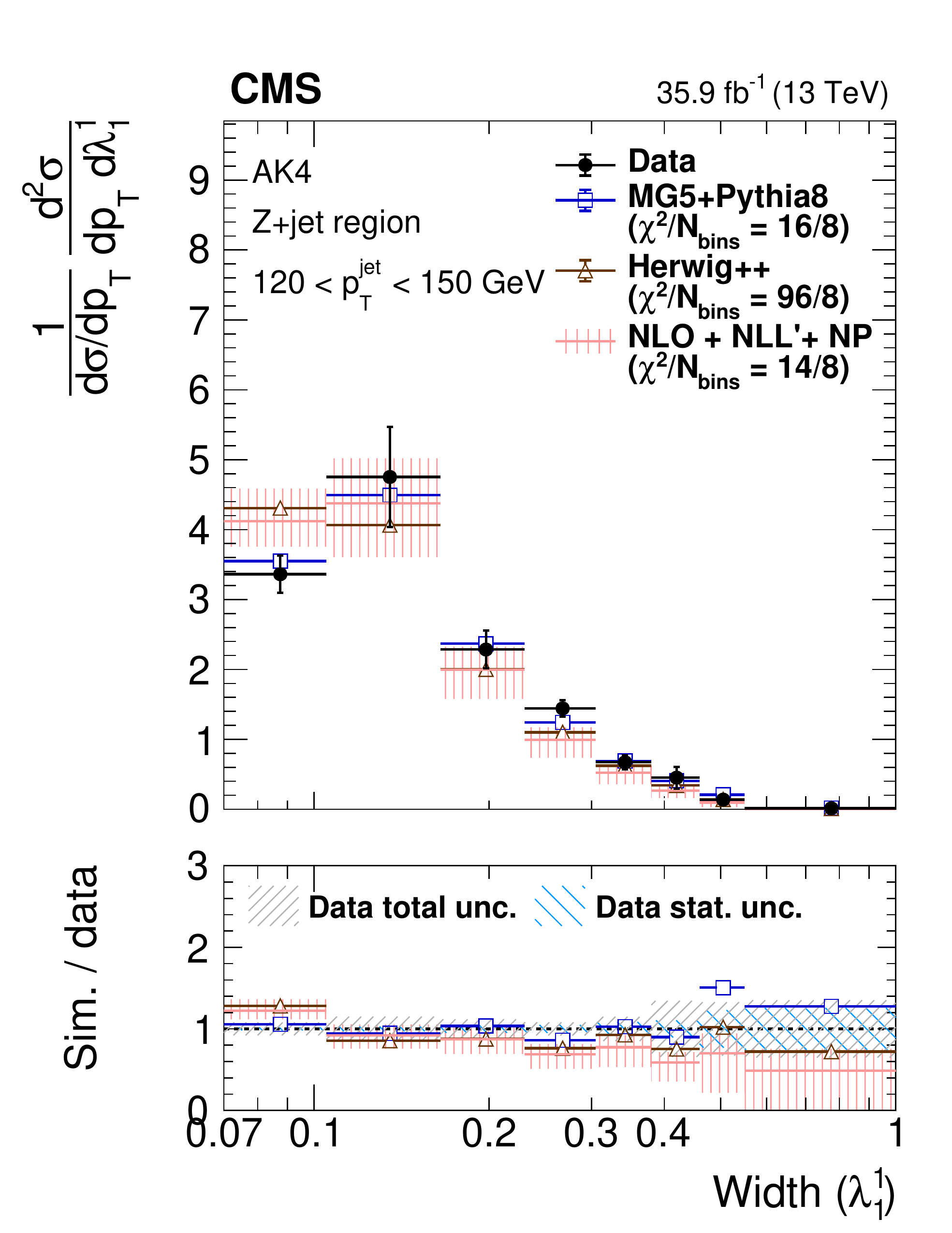}
    \includegraphics[width=0.49\textwidth]{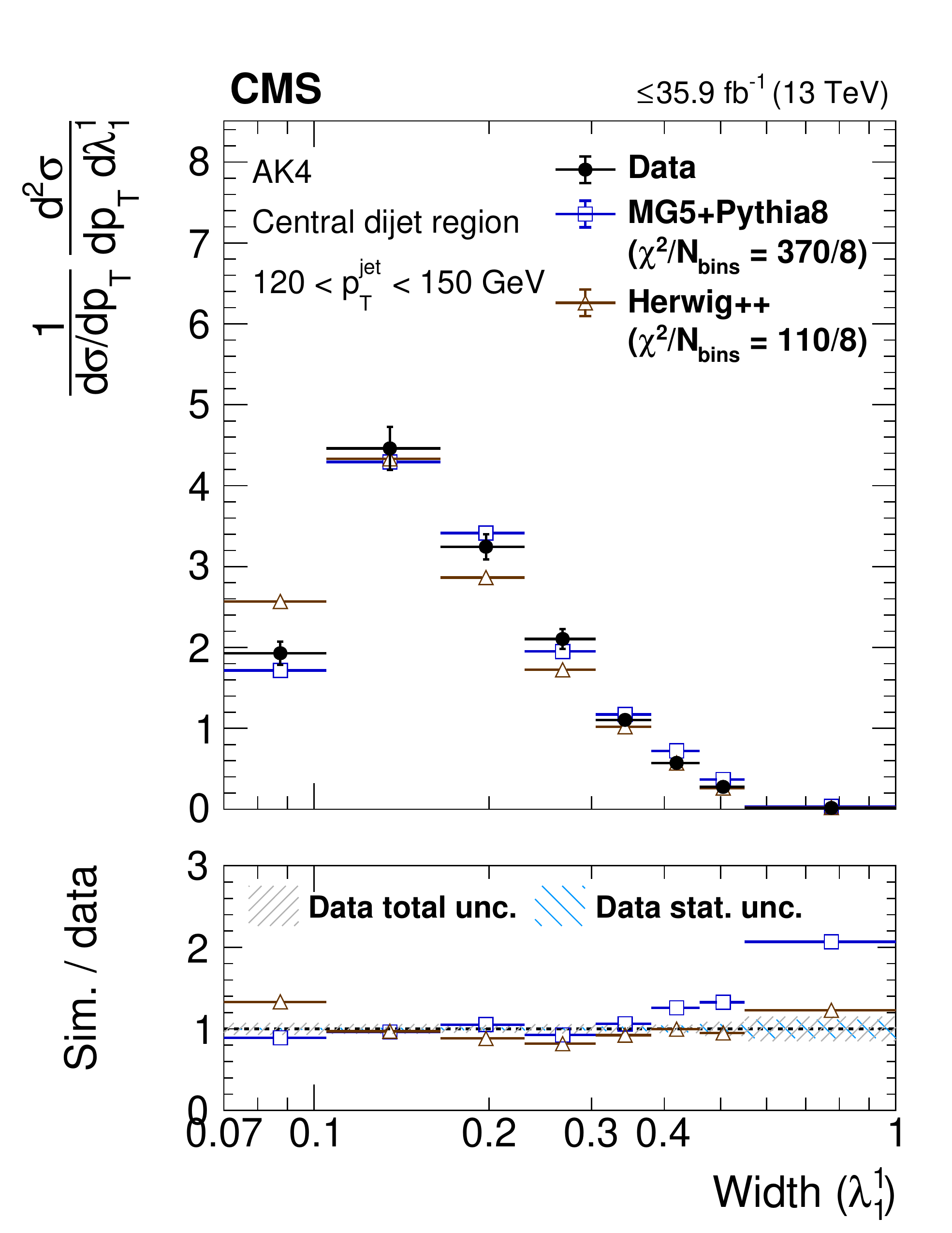}
    \caption{Particle-level distributions of (upper) ungroomed thrust (\thrust) and (lower) ungroomed width (\width) in $120<\pt<150$\GeV in the \Zjet region (left) and central dijet region (right).
    The error bars on the data correspond to the total uncertainties.
    For the NLO+NLL'+NP prediction, the theory uncertainty is displayed as a red hashed band.
    The coarse-grained blue hashed region in the ratio plot indicates the statistical uncertainty of the data, and the fine-grained grey hashed region represents the total uncertainty.
    The lowest bin extends down to $\lambdakb>=0$.
    }
    \label{fig:UnfoldedSpectra3}
\end{figure}

\begin{figure}[!htb]
    \centering
    \includegraphics[width=0.49\textwidth]{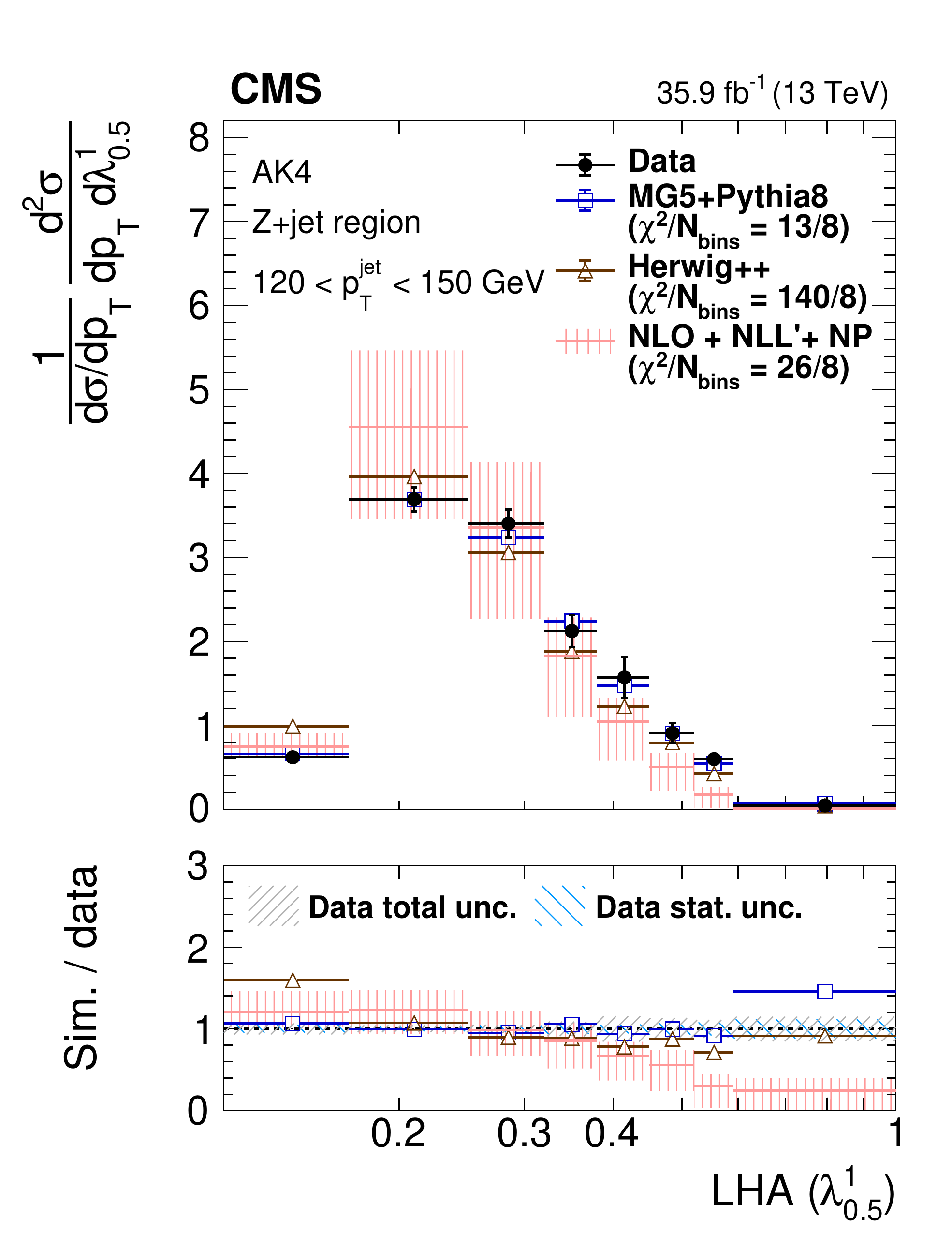}
    \includegraphics[width=0.49\textwidth]{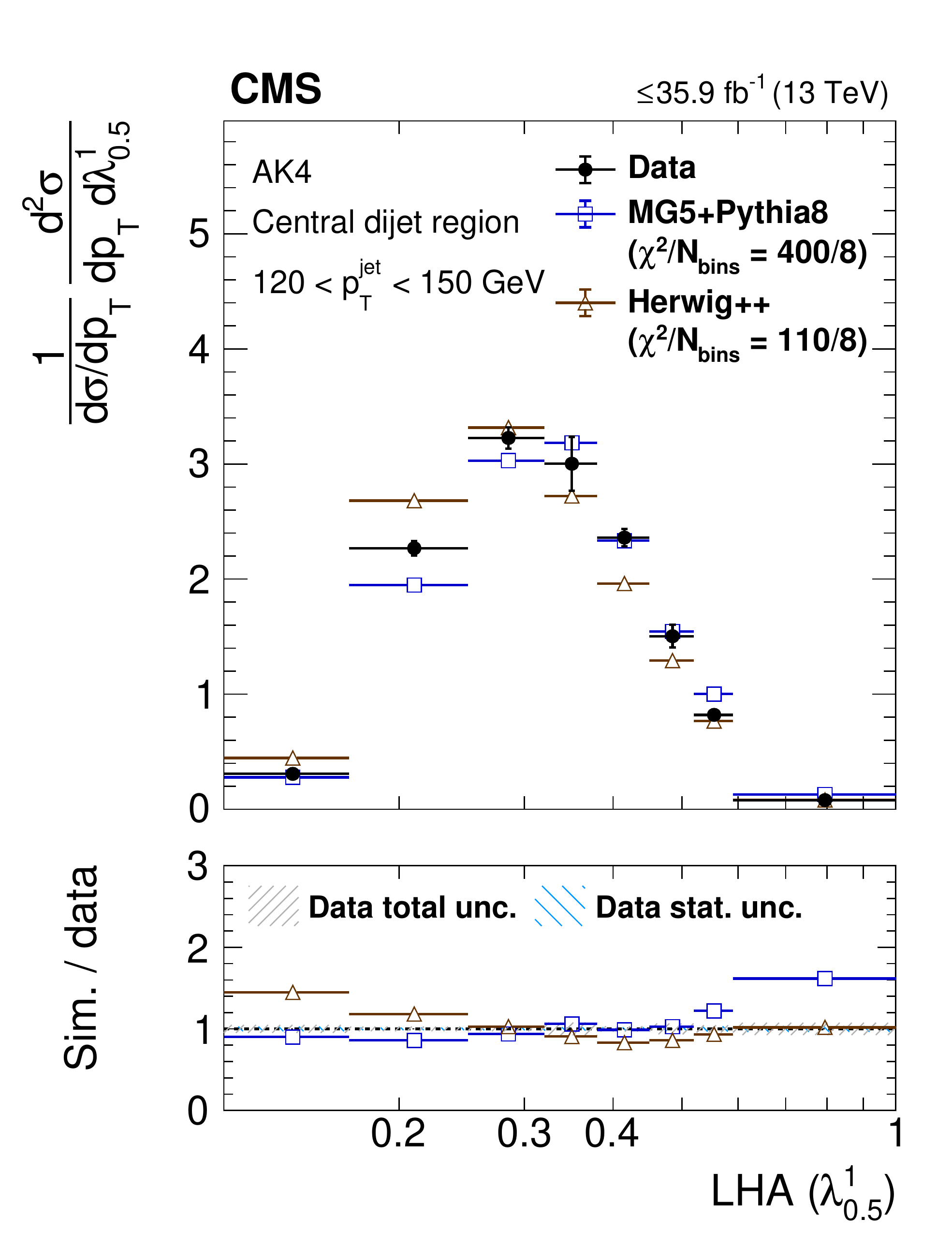}
    \caption{Particle-level distributions of ungroomed LHA (\lha) in $120<\pt<150$\GeV in the \Zjet region (left) and central dijet region (right).
    The error bars on the data correspond to the total uncertainties.
    For the NLO+NLL'+NP prediction, the theory uncertainty is displayed as a red hashed band.
    The coarse-grained blue hashed region in the ratio plot indicates the statistical uncertainty of the data, and the fine-grained grey hashed region represents the total uncertainty.
    The lowest bin extends down to $\lambdakb>=0$.
    }
    \label{fig:UnfoldedSpectra1}
\end{figure}

Experimental data distributions unfolded to particle level in the \Zjet and central dijet regions are presented in Figs.~\ref{fig:UnfoldedSpectra4}--\ref{fig:UnfoldedSpectra1} for the ungroomed substructure observables of AK4 jets with $120<\pt<150\GeV$.
The data are compared with {\mgpy} and \HERWIGpp predictions at particle level.
For the IRC-safe observables (LHA, width, and thrust) in the \Zjet region, the data are also compared with predictions from Ref.~\cite{Caletti:2021oor} based on analytic resummation of large logarithms at next-to-leading logarithmic accuracy (NLL), matched to the exact NLO prediction, plus nonperturbative (NP) corrections derived from Sherpa, labeled NLO+NLL'+NP, where the superscript ' of NLL indicates that a matching procedure keeping track of the jet flavour is used.
The uncertainty band of the NLO+NLL'+NP prediction includes variations of factorization, renormalization and resummation scales as well as the model for NP corrections.
For intermediate values of \lambdakb the prediction has small contributions from NP effects, which dominate at very low and high values of \lambdakb.

As a simple quantitative assessment of agreement of each prediction with CMS data, values of $\chi^2/N_{\text{bins}}$ are quoted, where $N_{\text{bins}}$ is the number of bins of the distribution.
The $\chi^2/N_{\text{bins}}$ is computed including the covariance between the measured cross sections including all statistical and experimental systematic uncertainties.
For the NLO+NLL'+NP prediction theoretical uncertainties are taken into account as well, whereas for the MC generator prediction no theoretical systematic uncertainties are considered.

The agreement of the {\mgpy} and \HERWIGpp simulations with the experimental data varies depending on the observable and sample.
In both \Zjet and central dijet regions, the {\mgpy} and \HERWIGpp predictions tend to envelop the experimental data distributions.
The significant difference between the {\mgpy} and \HERWIGpp predictions reflects the uncertainty in the prediction from MC event generators.
These two predictions differ in the modelling of perturbative (matrix element, parton shower) and nonperturbative (hadronization) effects.
In the \Zjet region, {\mgpy} provides the best description of all angularities.
\HERWIGpp predicts smaller values than {\mgpy} for all observables, except \pTD.
In the central dijet region, the agreement of both generators is worse than in the \Zjet region.
Here, \HERWIGpp shows a slightly better description of the experimental data than {\mgpy}, except for multiplicity and \pTD.

The NLO+NLL'+NP prediction describes the thrust well, with a $\chi^2/N_{\text{bins}}$ below unity.
The description of the width distribution is slightly worse, in particular at very low and high values of width, where NP effects contribute.
The LHA shows a significant disagreement, indicating the need for further investigation of this prediction.
These observations are consistent with the conclusions from Ref.~\cite{Caletti:2021oor} where reasonable control of the thrust and width was obtained, although the LHA was not well under control.
In the following, we focus on the LHA distribution where data may guide investigations.

\begin{figure}[!htp]
    \centering
    \includegraphics[width=0.49\textwidth]{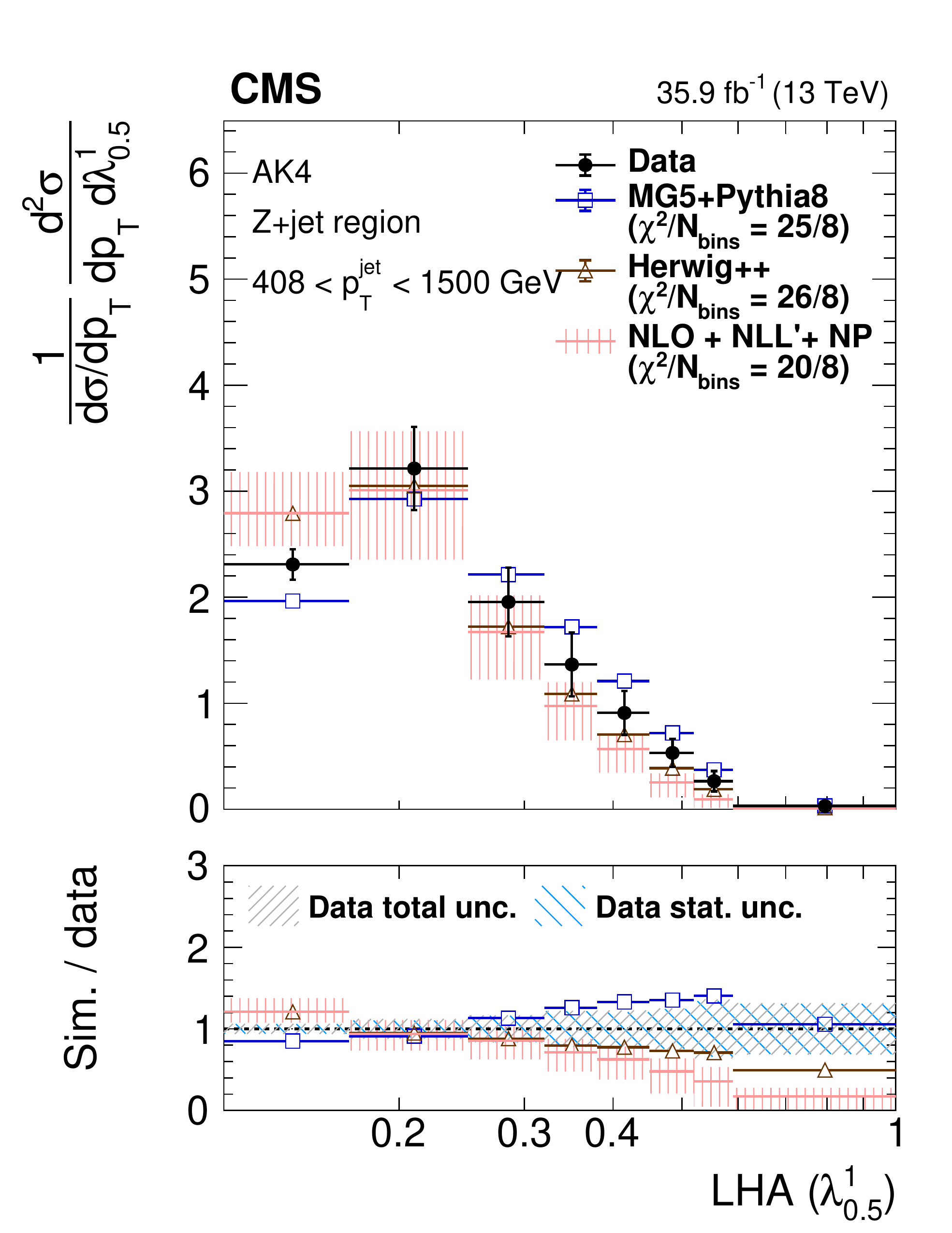}
    \includegraphics[width=0.49\textwidth]{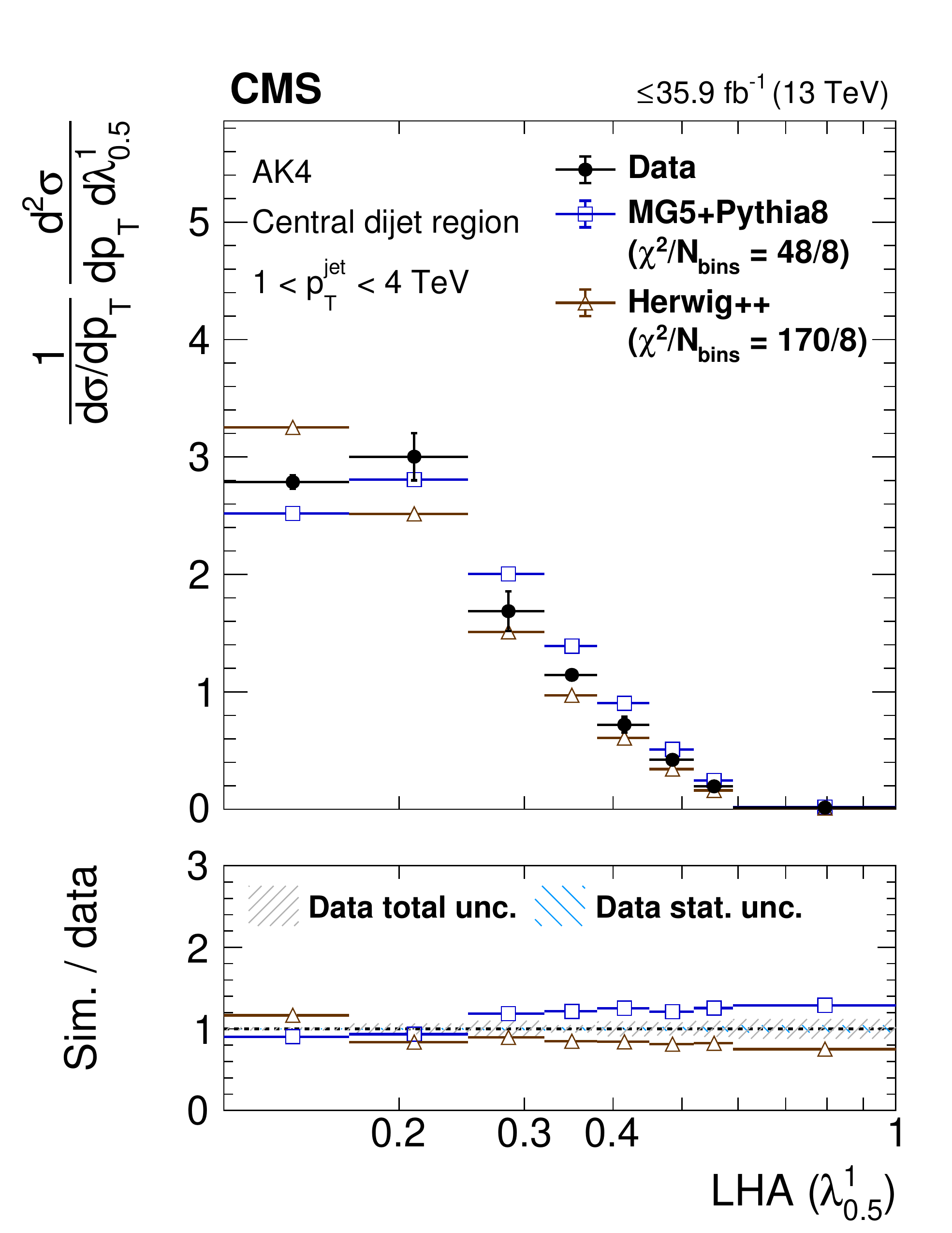} \\
    \includegraphics[width=0.49\textwidth]{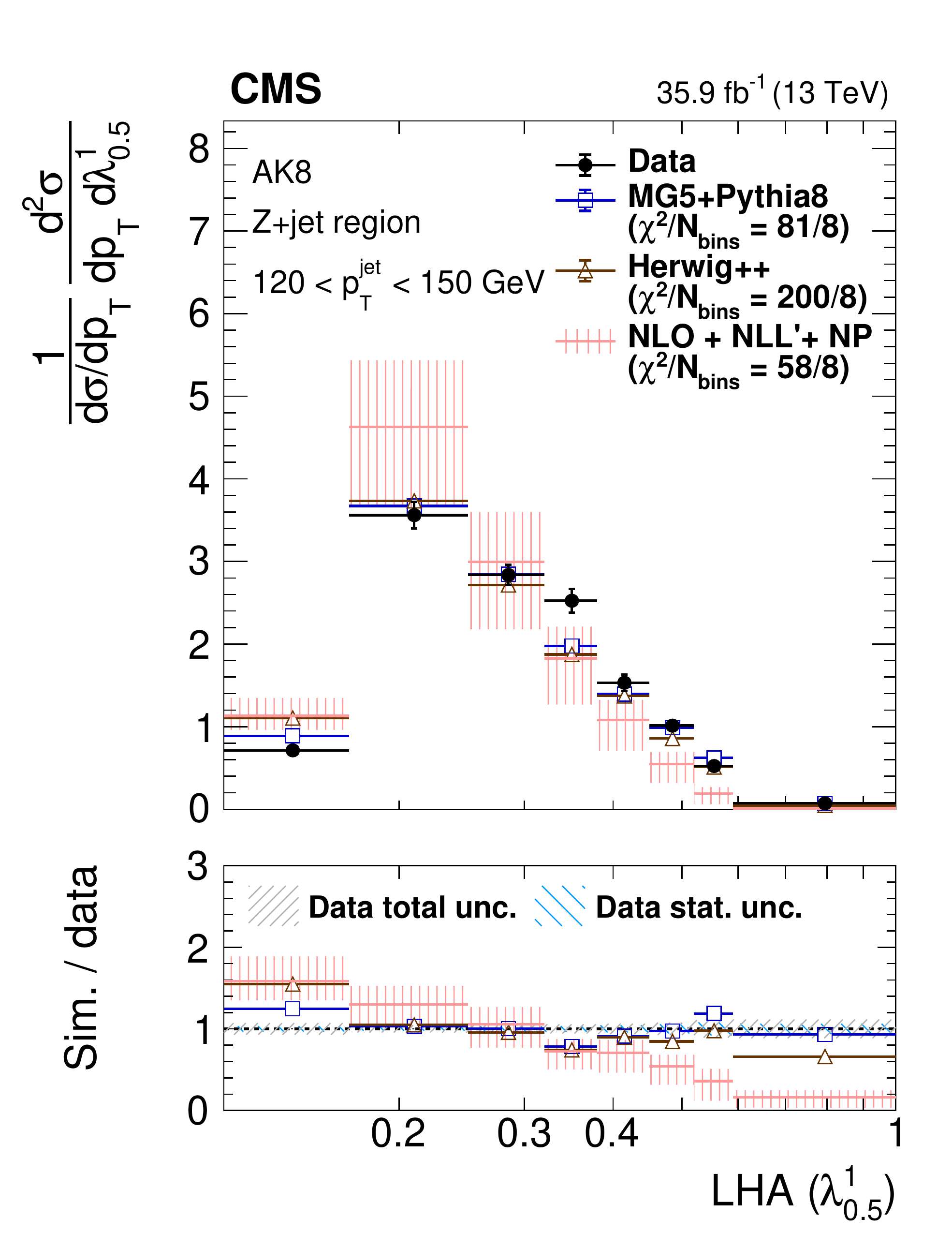}
    \includegraphics[width=0.49\textwidth]{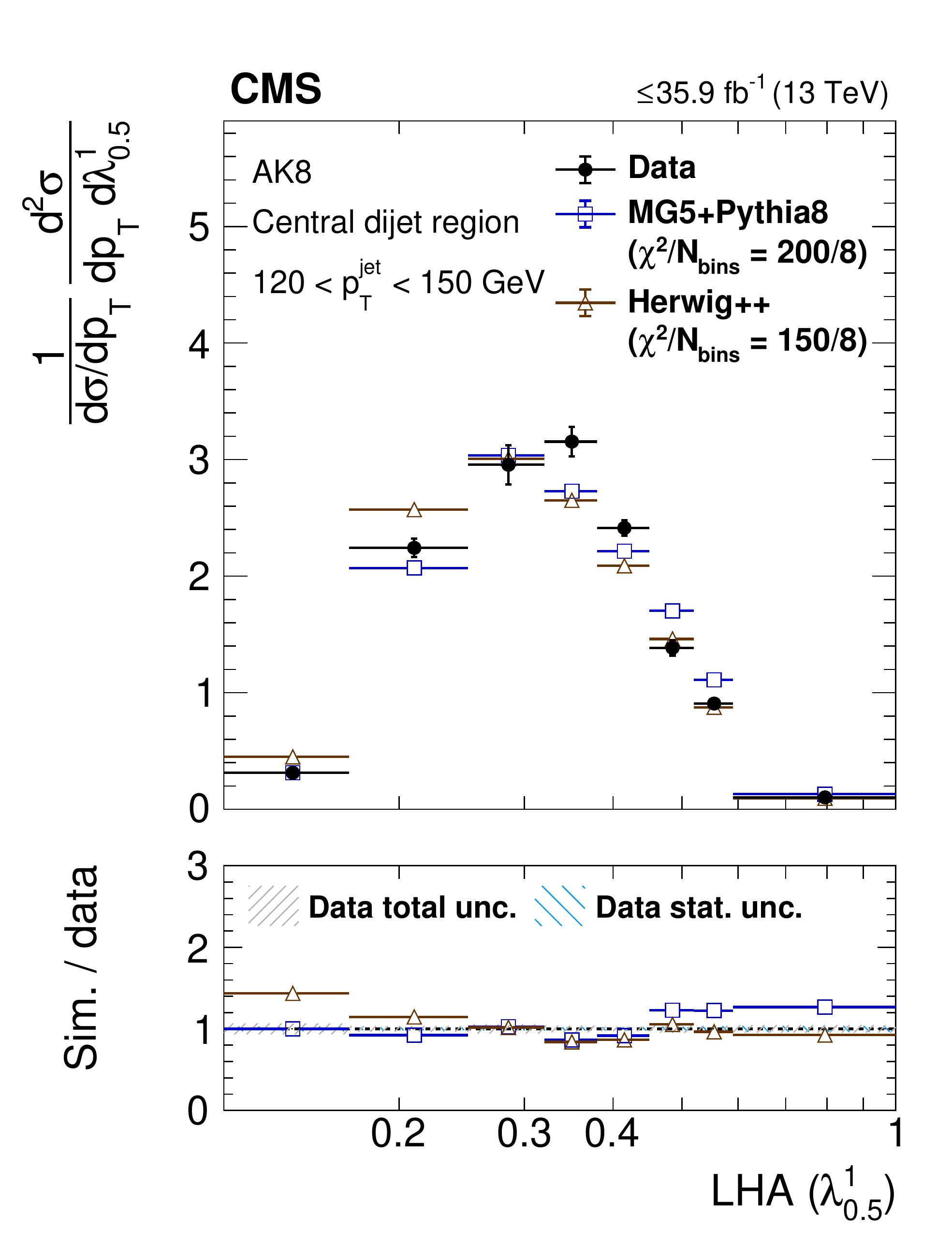}
    \caption{Particle-level distributions of (upper) ungroomed AK4 LHA (\lha) in $408<\pt<1500\GeV$ in the \Zjet region (left) and in $1<\pt<4\TeV$ in the central dijet region (right) and (lower) ungroomed AK8 LHA (\lha) in AK8 $120<\pt<150\GeV$ in the \Zjet region (left) and central dijet region (right).
    The error bars on the data correspond to the total uncertainties.
    For the NLO+NLL'+NP prediction, the theory uncertainty is displayed as a red hashed band.
    The coarse-grained blue hashed region in the ratio plot indicates the statistical uncertainty of the data, and the fine-grained grey hashed region represents the total uncertainty.
    The lowest bin extends down to $\lambdakb>=0$.
    }
    \label{fig:UnfoldedSpectra6}
\end{figure}

\begin{figure}[!htp]
    \centering
    \includegraphics[width=0.49\textwidth]{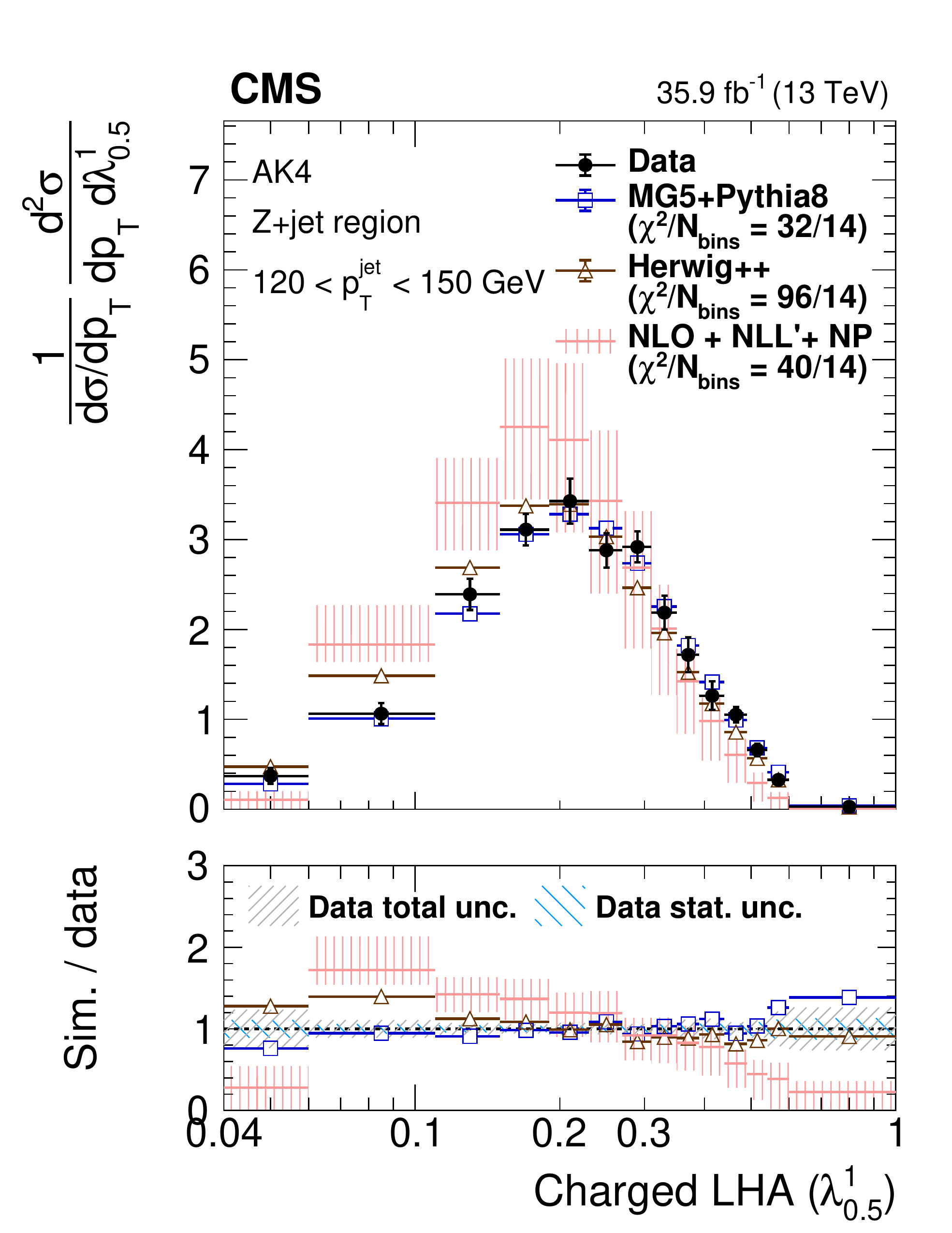}
    \includegraphics[width=0.49\textwidth]{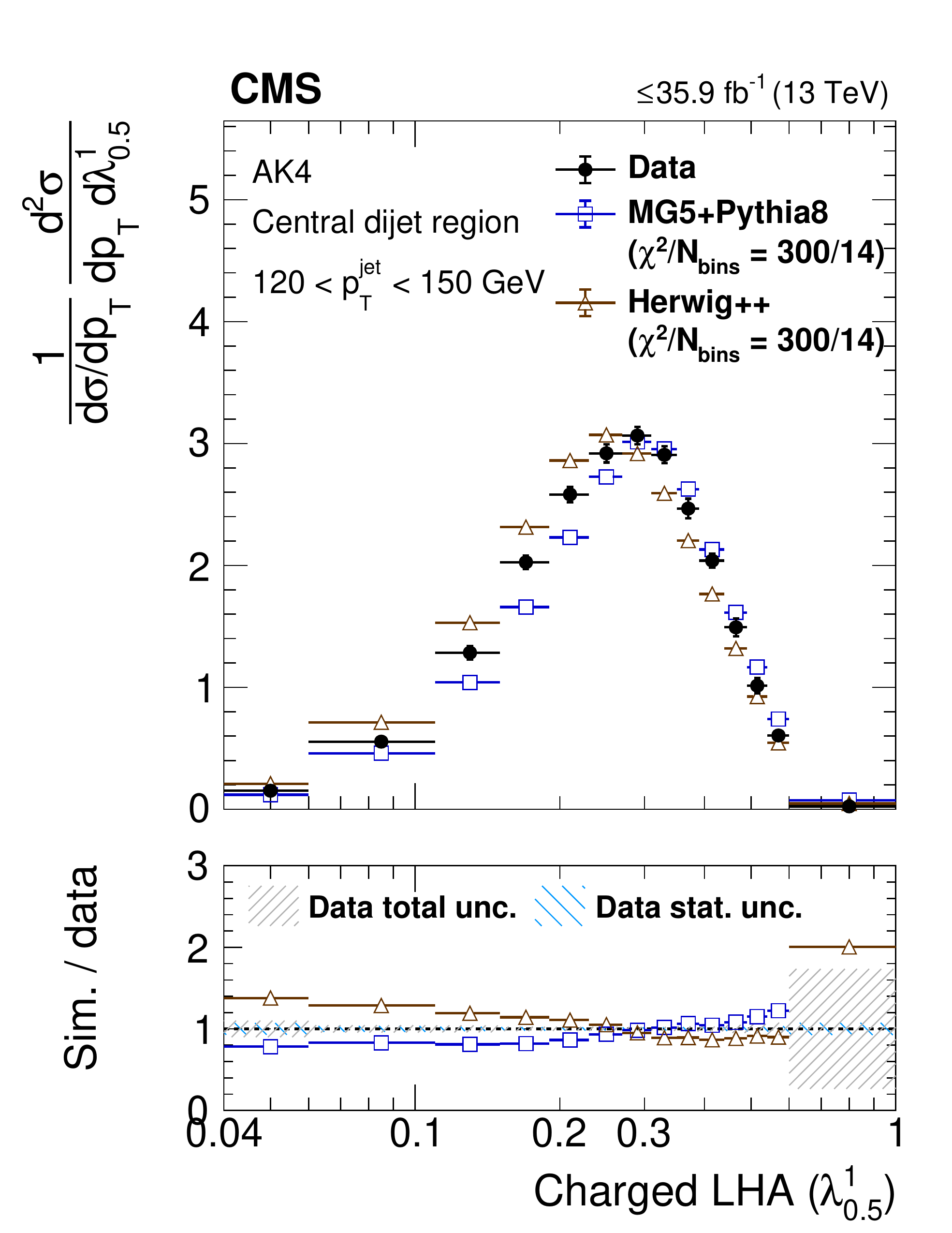} \\
    \includegraphics[width=0.49\textwidth]{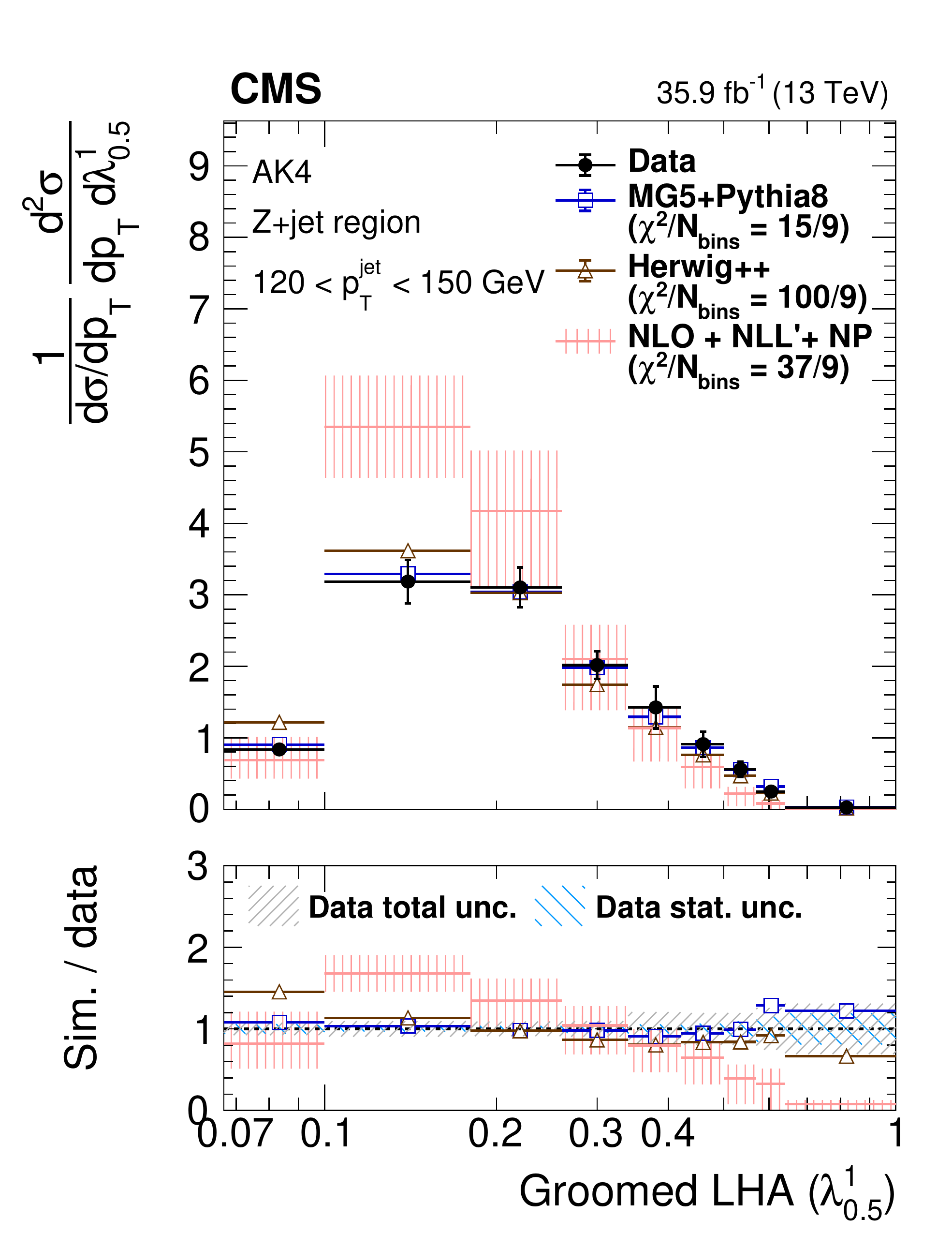}
    \includegraphics[width=0.49\textwidth]{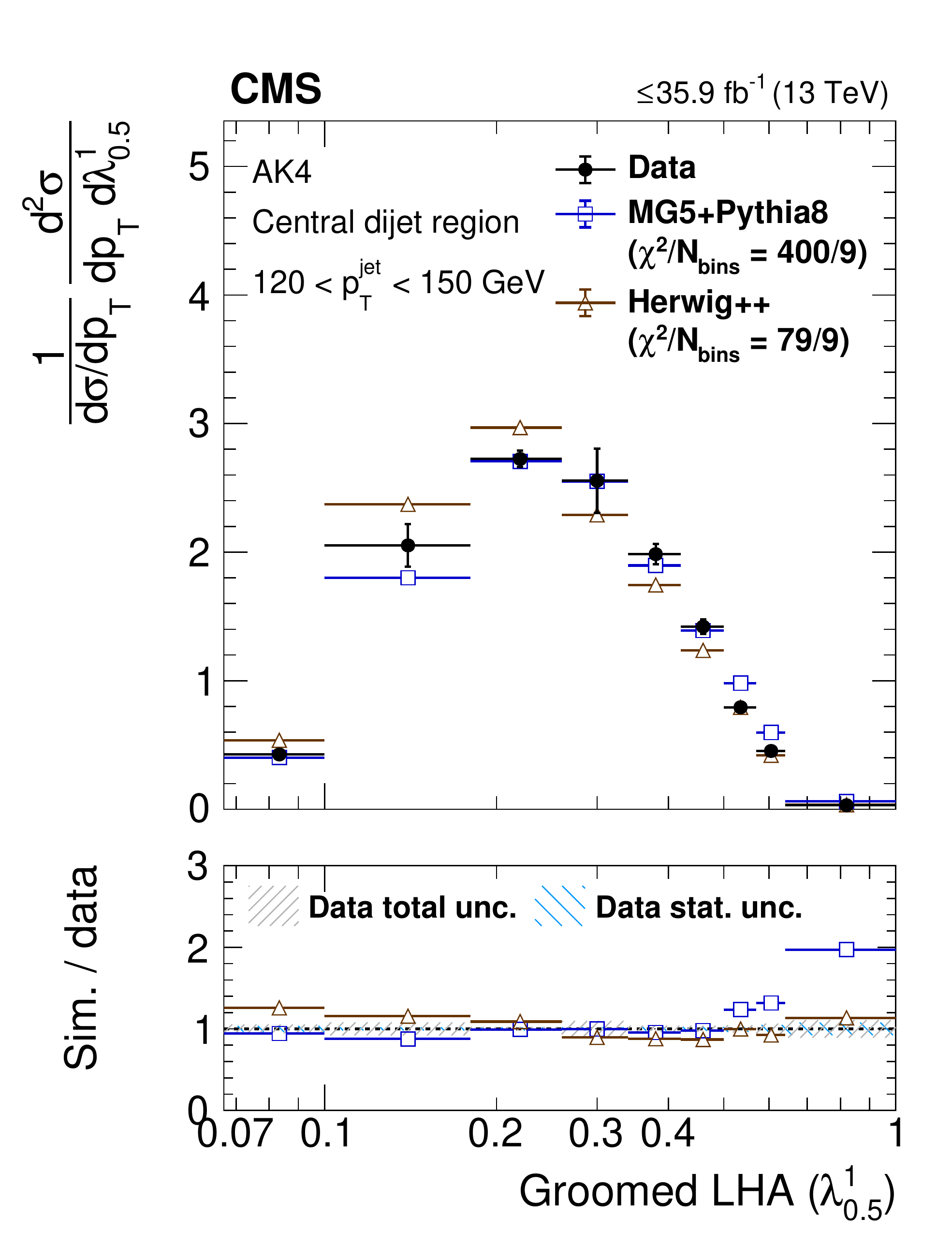}
    \caption{Particle-level distributions of (upper) ungroomed charged AK4 LHA (\lha) and (lower) groomed AK4 LHA (\lha) in $120<\pt<150\GeV$ in the \Zjet region (left) and central dijet region (right).
    The error bars on the data correspond to the total uncertainties.
    For the NLO+NLL'+NP prediction, the theory uncertainty is displayed as a red hashed band.
    The coarse-grained blue hashed region in the ratio plot indicates the statistical uncertainty of the data, and the fine-grained grey hashed region represents the total uncertainty.
    The lowest bin extends down to $\lambdakb>=0$.
    }
    \label{fig:UnfoldedSpectra8}
\end{figure}

Figures~\ref{fig:UnfoldedSpectra6}--\ref{fig:UnfoldedSpectra8} show multiple variants of LHA.
Based on the $\chi^2/N_{\text{bins}}$, the level of data-to-simulation agreement of {\mgpy} for AK4 jets with $408<\pt<1500\GeV$ in the \Zjet region (Fig.~\ref{fig:UnfoldedSpectra6}, upper left) is worse than in the low-\pt \Zjet region (Fig.~\ref{fig:UnfoldedSpectra1}, left).
The level of data-to-simulation agreement of {\mgpy} for $1<\pt<4\TeV$ in the central dijet region (enriched in quark jets, Fig.~\ref{fig:UnfoldedSpectra6}, upper right) is better than in the low-\pt central dijet region (enriched in gluon jets, Fig.~\ref{fig:UnfoldedSpectra1}, right).
For \HERWIGpp, the level of agreement changes in opposite directions going from low-\pt to high-\pt.
This is consistent with the hypothesis that the level of agreement is related to the expected gluon fraction in the sample.
The NLO+NLL'+NP prediction yields a similar level of agreement at low-\pt and high-\pt.

The LHA distribution for AK8 jets with $120<\pt<150\GeV$ (Fig.~\ref{fig:UnfoldedSpectra6} lower) is similar to that for AK4 jets.
Its level of agreement between data and the {\mgpy} and NLO+NLL'+NP predictions is worse than for AK4 jets, although the level of agreement remains more similar for \HERWIGpp.

Compared to LHA from charged+neutral constituents, the charged LHA distribution for ungroomed AK4 jets with $120<\pt<150\GeV$ (Fig.~\ref{fig:UnfoldedSpectra8} upper) is binned more finely taking advantage of the good track resolution, resolving LHA values well below 0.1.
The data-to-theory agreement is similar in magnitude and shape above 0.1, though the $\chi^2/N_{\text{bins}}$ is worse, suggesting that the majority of the underlying difference to experimental data in the theory prediction may be probed using a charged observable.
This measurement is consistent with conclusions in Ref.~\cite{Caletti:2021oor}, where the NP corrections were very similar between charged+neutral and charged observables.

The LHA distribution for groomed AK4 jets with $120<\pt<150\GeV$ (Fig.~\ref{fig:UnfoldedSpectra8} lower) is wider than for ungroomed jets.
The data-to-theory agreement for LHA in groomed jets is similar compared with that in ungroomed jets.
Although grooming is expected to reduce the influence from pileup, underlying event, and initial-state radiation, which are all difficult to model, we find no significant improvement in the description of LHA with grooming.
This measurement is consistent with conclusions in Ref.~\cite{Caletti:2021oor}, where groomed LHA has a large remaining contribution from NP effects.

\begin{figure}[!htp]
    \centering
    \includegraphics[width=0.49\textwidth]{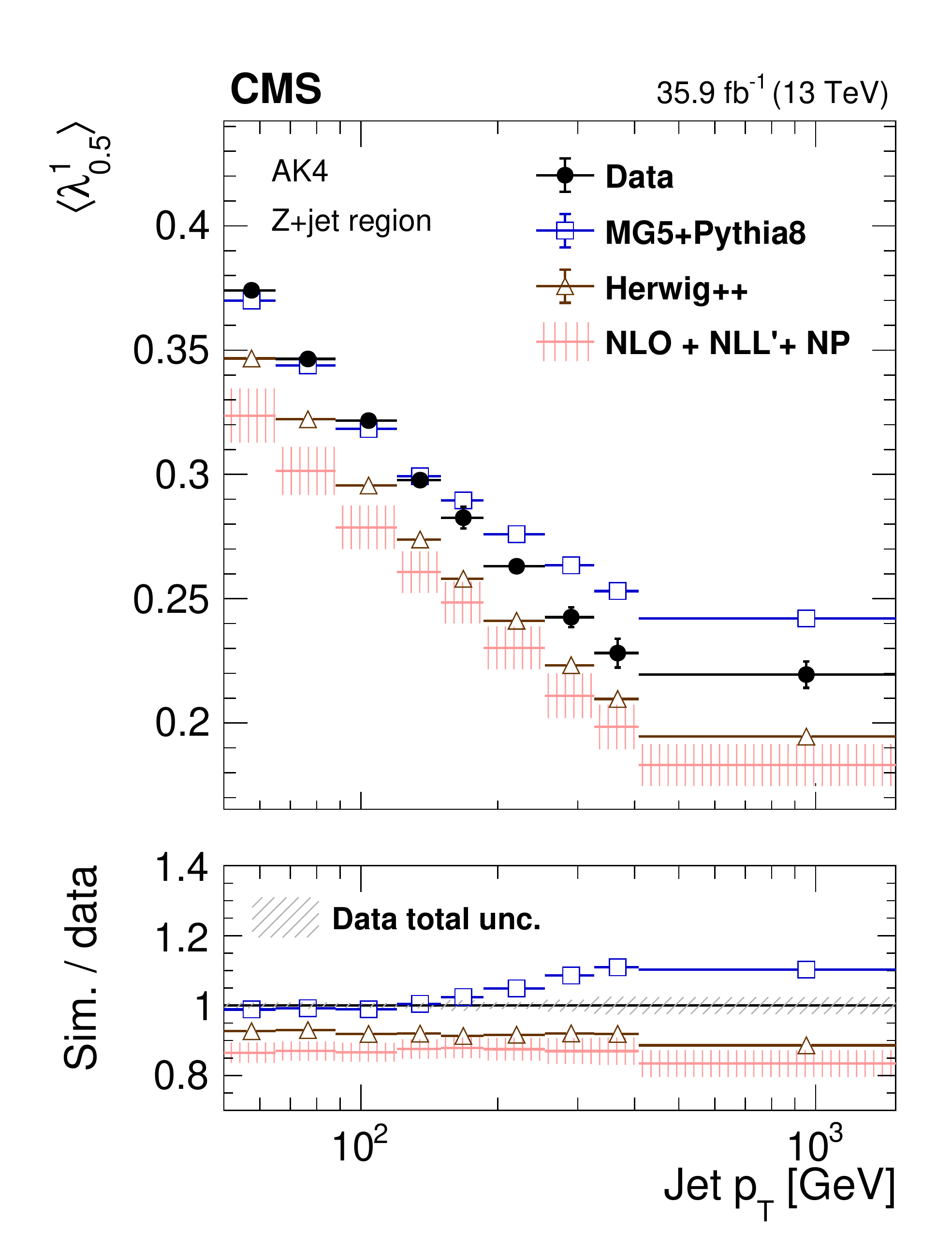}
    \includegraphics[width=0.49\textwidth]{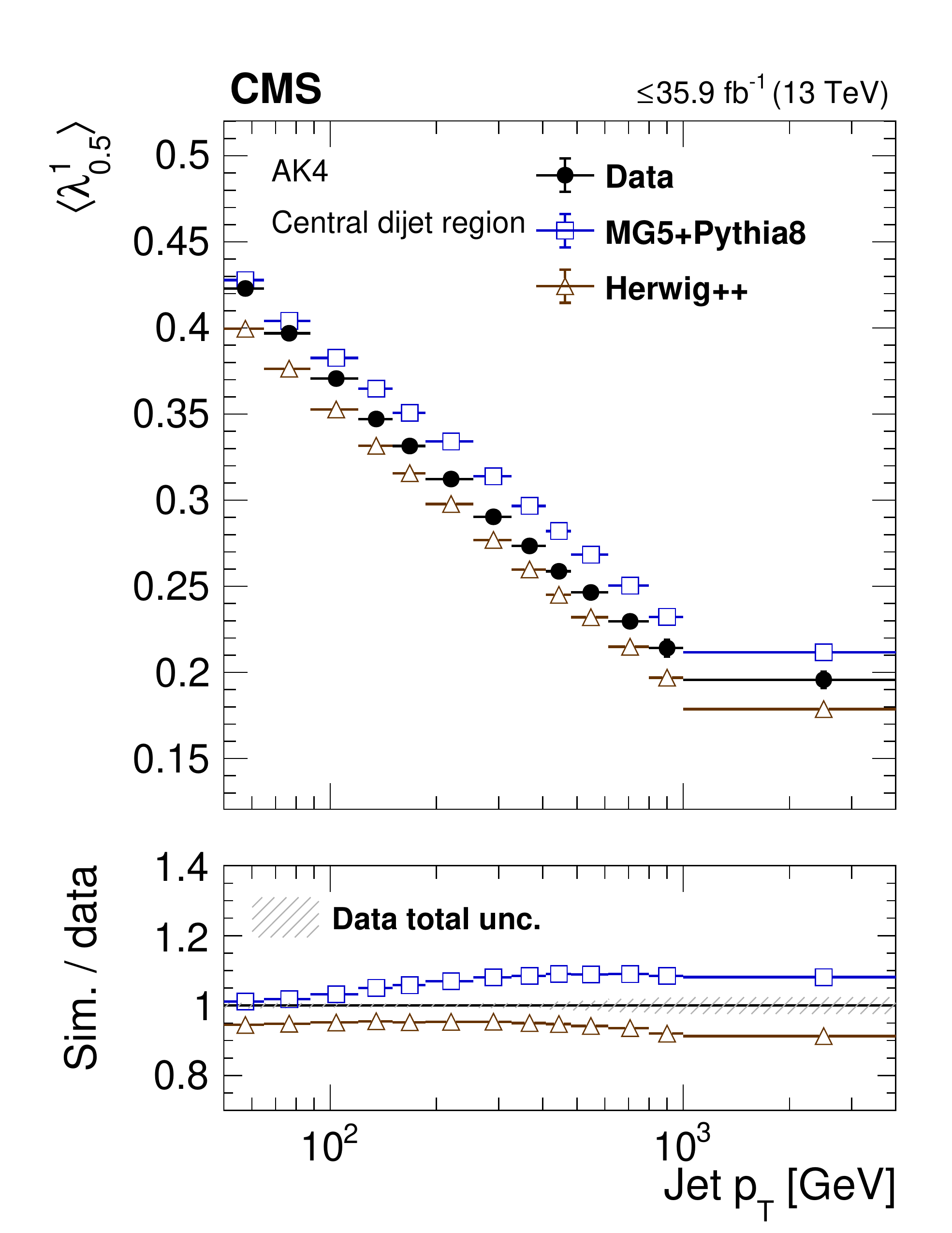}\\
    \includegraphics[width=0.49\textwidth]{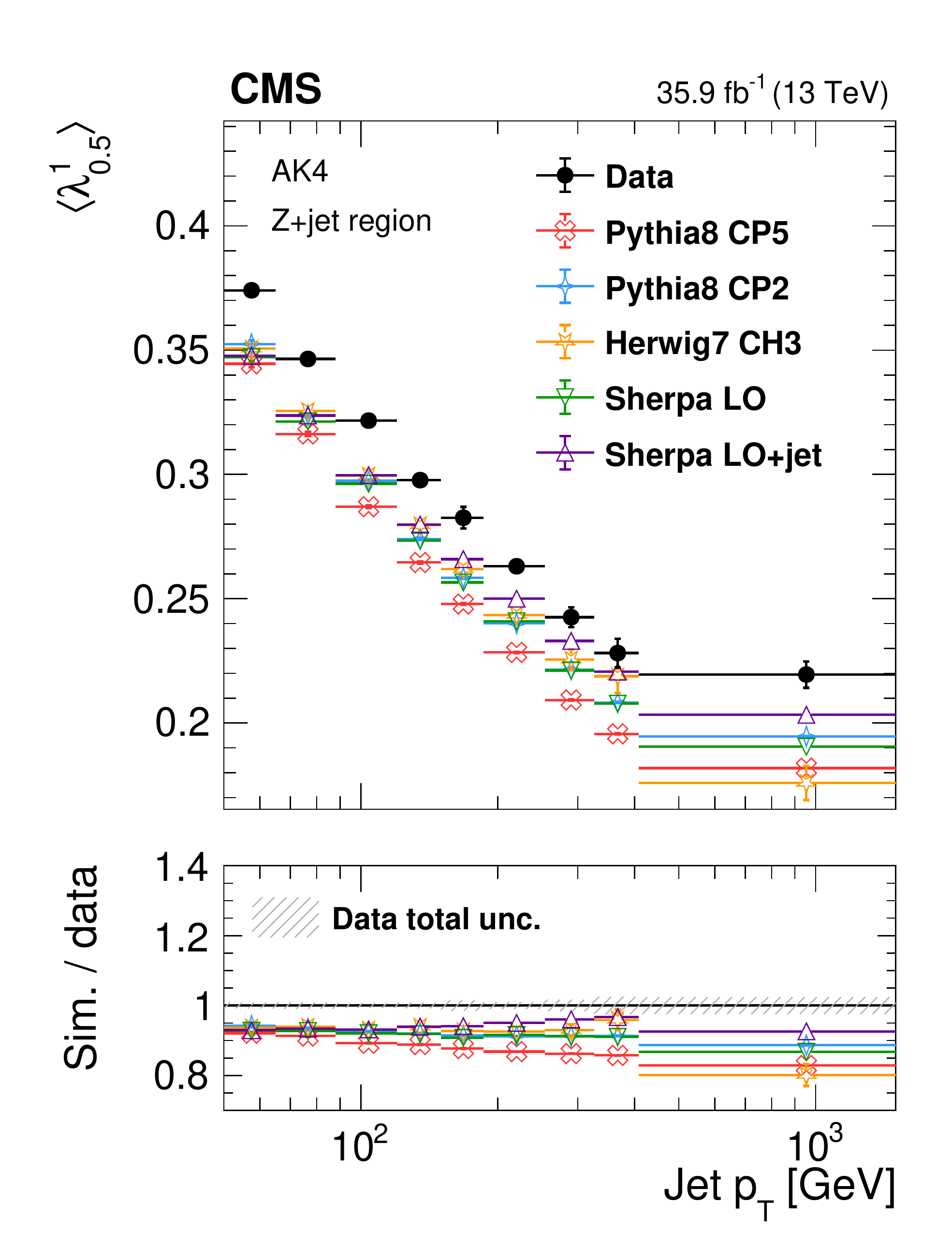}
    \includegraphics[width=0.49\textwidth]{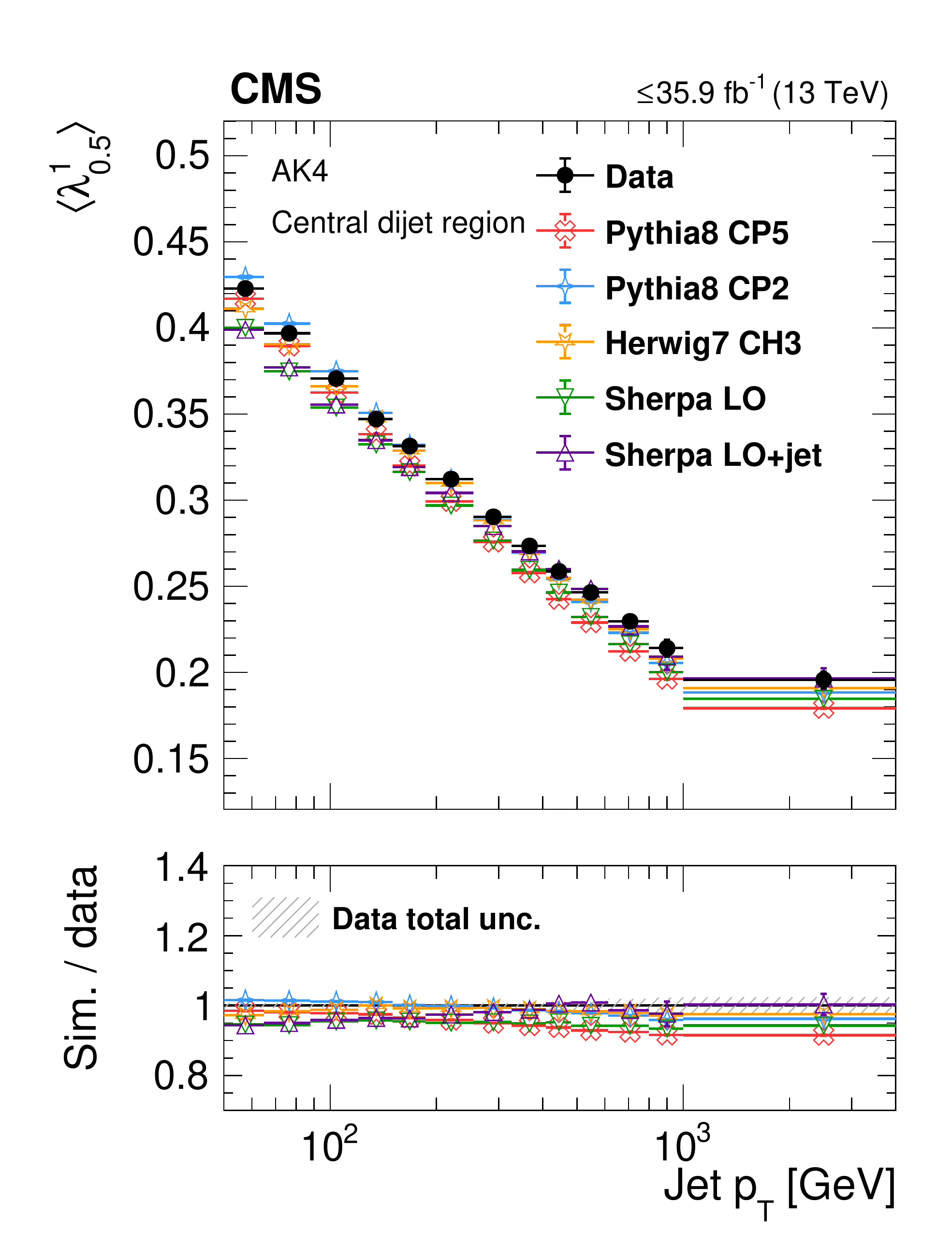}
    \caption{Mean value of ungroomed LHA (\lha) for AK4 jets as a function of \pt in the \Zjet (left) and central dijet region (right) regions.
    The upper and lower plots show the same data distribution compared with different generator predictions.
    The error bars on the data and the hashed region in the ratio plot correspond to the total uncertainties of the experimental data.
    The error bars on the simulation correspond to the statistical uncertainties.}
    \label{fig:pTsummary}
\end{figure}

To study the behaviour of the measured jet substructure observables in the dimensions summarized in Table~\ref{tab:observables}, we focus on the mean value of each substructure distribution.
Figure~\ref{fig:pTsummary} shows the measured mean of ungroomed LHA (\lha) for AK4 jets as a function of jet \pt, including also predictions from more recent generators.
The mean decreases with the jet \pt in both the \Zjet and central dijet regions, as a result of constituents being located closer to the jet axis due to the larger Lorentz boost at higher \pt.
This trend is displayed by all generators.
The mean in the \Zjet region is well-described at low \pt and overestimated at high \pt by {\mgpy}, whereas all other predictions significantly underestimate it across the entire \pt range.
In the central dijet sample, {\mgpy} and \HERWIGpp generator predictions deviate significantly from the measurement.
Although {\mgpy} predicts a larger mean than that measured in the experimental data, \HERWIGpp underestimates it across the whole \pt range.
The \HERWIG{}7 CH3 and \PYTHIA{8} CP2 simulations provide the best description, followed by \PYTHIA{8} CP5 and \SHERPA.
One can infer from this observation that the data-to-simulation agreement depends on how generators model the difference between quark and gluon jets, consistent with previous measurements of observables related to jet fragmentation~\cite{Chatrchyan:2012mec}.

\begin{figure}[!htp]
    \centering
    \includegraphics[width=0.96\textwidth]{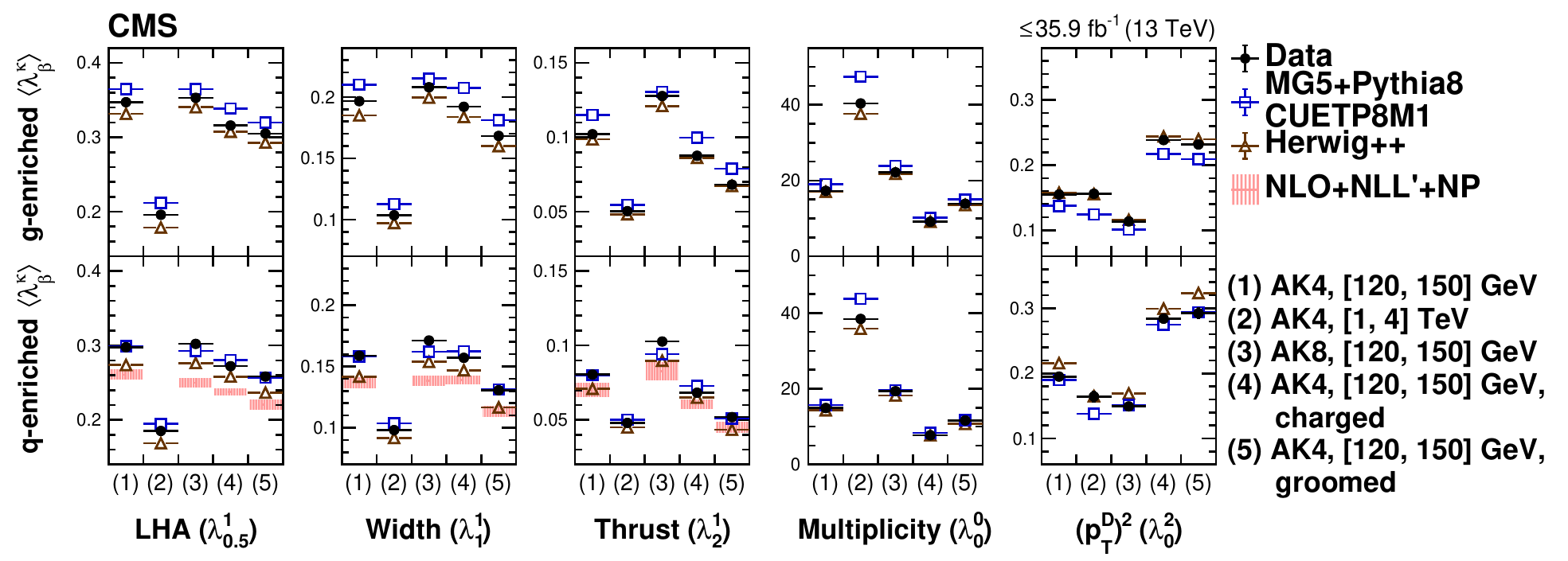} \\
    \includegraphics[width=0.96\textwidth]{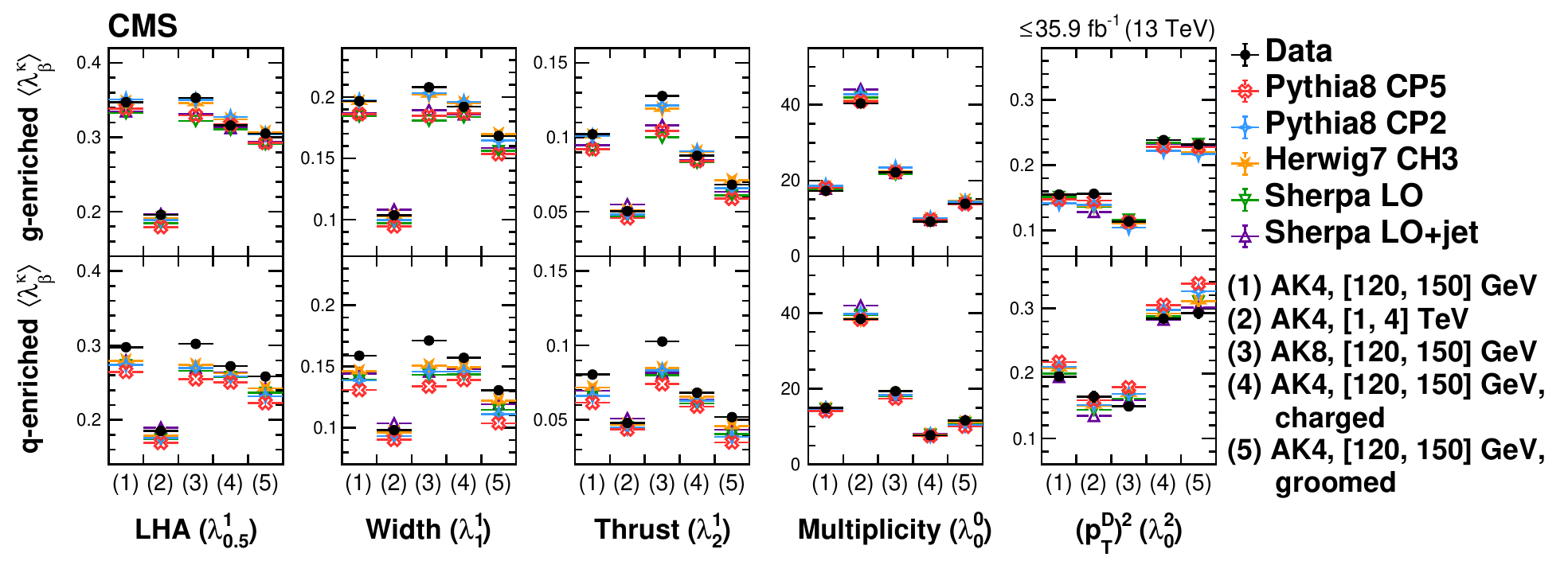}
    \caption{Mean value of substructure observables in regions with gluon-enriched and quark-enriched jets, for the following configurations:
    (1) ungroomed AK4 $120<\pt<150\GeV$, (2) ungroomed AK4 $1<\pt<4\TeV$, (3) ungroomed AK8 $120<\pt<150\GeV$, (4) ungroomed charged AK4 $120<\pt<150\GeV$, and (5)  groomed AK4 $120<\pt<150\GeV$; shown for each of the observables LHA (\lha), width (\width), thrust (\thrust), multiplicity (\multi), and \pTD (\ptd). The central jet in the dijet region is used for the gluon-enriched sample, whereas for the quark-enriched sample the jet in the \Zjet region is used for $120<\pt<150\GeV$, and the forward jet in the dijet region is used for $1<\pt<4\TeV$.
    The upper and lower plots show the same data distribution compared with different generator predictions.
    The error bars on the data correspond to the total uncertainties.
    The error bars on the simulation correspond to the statistical uncertainties.}
    \label{fig:grantSummary1}
\end{figure}

Figure~\ref{fig:grantSummary1} summarizes the behaviour of jet substructure observables across all dimensions under study using the mean of each distribution.
The upper plot in Fig.~\ref{fig:grantSummary1} shows the results from CMS data and the {\mgpy} and \HERWIGpp simulations.
Across most observables and dimensions, the measurements are enveloped by the two generator predictions.
The behaviour as a function of \pt and jet radius parameter as well as the behaviour of charged and groomed observables discussed previously for LHA holds also for the other substructure observables.

Figure~\ref{fig:grantSummary1} (lower) compares the same measured data to predictions from more recent generators.
In the gluon-enriched sample, \HERWIG{}7 CH3, \PYTHIA{8} CP5, \PYTHIA{8} CP2, and \SHERPA generally provide a better description than either \HERWIGpp or {\mgpy}.
In the quark-enriched sample, {\mgpy} provides the best description, followed by
\HERWIG{}7, \PYTHIA{8} CP2, \SHERPA, \HERWIGpp, and \PYTHIA{8} CP5.
Thus, the previous observations in Fig.~\ref{fig:grantSummary1} (upper) of more accurate modelling of quark jet substructure, and less accurate modelling of gluon jet substructure, does not necessarily hold for more recent generators and tunes under study.
Improved modelling of gluon jets at the cost of poorer modelling of quark jets is observed.
In addition to the more recent generator versions and tunes used for \HERWIG and \PYTHIA{8}, an important difference among the predictions is the choice of $\alpS(m_{\PZ})$ in the final state shower, to which the jet substructure modelling is sensitive~\cite{Sirunyan:2018asm}.
The \PYTHIA{8} CP2 prediction using $\alpS(m_{\PZ})=0.130$ predicts systematically larger values of LHA, width, thrust, and multiplicity, and smaller value of \pTD, for both quarks and gluons than the \PYTHIA{8} CP5 prediction using $\alpS(m_{\PZ})=0.118$.

\begin{figure}[!htb]
    \centering
    \includegraphics[width=0.96\textwidth]{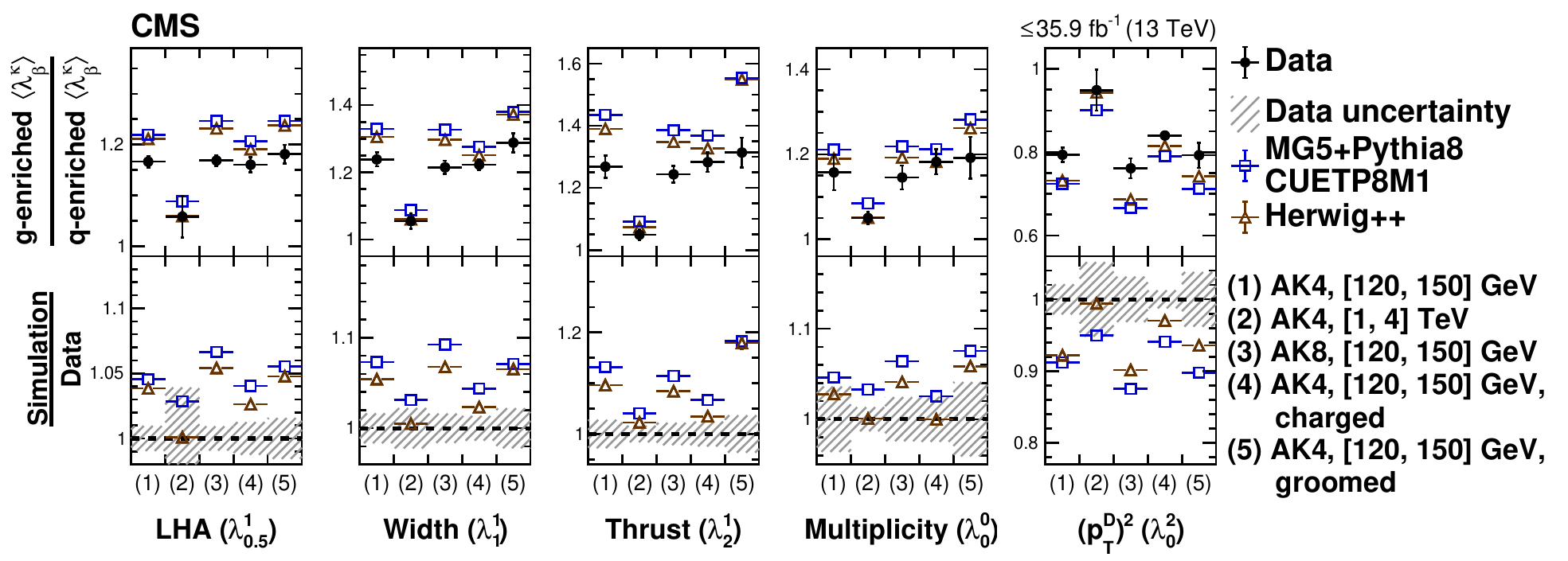} \\
    \includegraphics[width=0.96\textwidth]{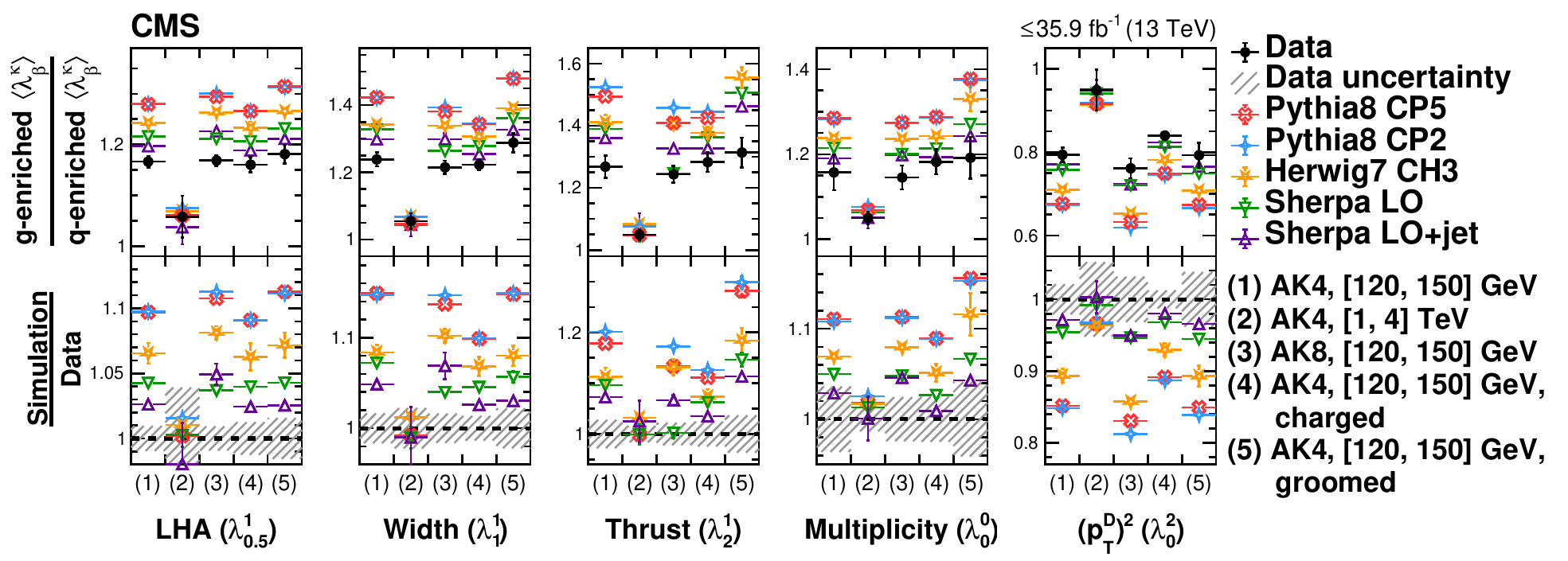}
    \caption{Ratio of the mean of substructure observables in regions with gluon- and quark-enriched jets, for the following configurations:
    (1) ungroomed AK4 $120<\pt<150\GeV$, (2) ungroomed AK4 $1<\pt<4\TeV$, (3) ungroomed AK8 $120<\pt<150\GeV$, (4) ungroomed charged AK4 $120<\pt<150\GeV$, and (5) groomed AK4 $120<\pt<150\GeV$; for the observables LHA (\lha), width (\width), thrust (\thrust), multiplicity (\multi), and \pTD (\ptd). The central jet in the dijet region is used for the gluon-enriched sample, whereas for the quark-enriched sample the jet in the \Zjet region is used for $120<\pt<150\GeV$, whereas the forward jet in the dijet region is used for $1<\pt<4\TeV$.
    The upper and lower plots show the same data distribution compared with different generator predictions.
    The error bars on the data correspond to the total uncertainties.
    The error bars on the simulation correspond to the statistical uncertainties.}
    \label{fig:grantSummary2}
\end{figure}

Figure~\ref{fig:grantSummary2} shows the ratio of the means in the gluon- and quark-enriched samples, illustrating how well the generators model both the differences between quarks and gluons, and the relative quark/gluon composition of the samples.
As previously mentioned, gluon jets exhibit, on average, larger values of LHA, width, thrust, and multiplicity, and smaller values of \pTD, than quark jets.
At low \pt, the ratio of the means in the central dijet and \Zjet regions is significantly larger than unity for LHA, width, thrust, and multiplicity, and significantly smaller for \pTD, indicating that these observables have significant separation power between quark and gluon jets.
Strikingly, all generators overestimate the difference between quark and gluon jets at low \pt when compared with experimental data.
At high \pt, however, the ratio of the means in the central and forward dijet regions is significantly closer to unity, and all generators give a reasonable description of the ratio that is consistent with that measured in data.
The gluon- and quark-enriched samples are each expected to have a more equal relative quark/gluon jet composition at this \pt (see Fig.~\ref{fig:flavFractions}).
The best overall data-to-simulation agreement for the ratio is achieved by \SHERPA, followed by \HERWIGpp, {\mgpy}, \HERWIG{}7 CH3, \PYTHIA{8} CP2, and \PYTHIA{8} CP5.
The \PYTHIA{8} CP2 and CP5 predictions are very similar, showing that the quark and gluon discrimination power is not governed by the choice of $\alpS(m_{\PZ})$ in the final state shower, but rather by the model used for jet fragmentation.
Comparing the behaviour of \SHERPA LO to LO+jet, the effect of the additional outgoing partons is an improvement in the description of the ratio of the means in the gluon- and quark-enriched samples.
Since the additional outgoing partons can affect the fraction of gluon jets in the signal regions, this demonstrates that the measurement of the ratio provides information not only about the differences between quarks and gluons, but also about the fraction of gluon jets.
A higher order prediction of the angularities in the dijet region, matching the precision of the NLO+NLL'+NP predictions in the \Zjet region, may yield better understanding of the source of this disagreement.

\section{Summary}
\label{sec:conclusions}

Measurements of distributions of generalized jet angularities in proton-proton collision data taken by the CMS detector at $\sqrt{s}=13\TeV$ in dijet and, for the first time, also in \Zjet topologies have been presented.
Whereas the dijet topology allows access to a sample of jets that predominantly originate from gluon fragmentation, the \Zjet topology yields a sample enriched in quark-initiated jets.
Five generalized angularities are measured to study different features in the modelling of jet substructure.
Three infrared- and collinear-safe angularities are particularly sensitive to perturbative emissions in jets, whereas the other two have larger contributions from nonperturbative effects.
For the first time, a measurement of angularities with different jet radii, both with and without the application of a grooming algorithm, was carried out to further discriminate between different features in the modelling.
Although a subset of these distributions was discussed in this paper, the full range of measurements is made public in HEPData record~\cite{hepdata}.
The measurements for quark and gluon jets yield values in between the predictions from the \MGvATNLO{}+\PYTHIA{8} and \HERWIGpp simulations.
The quality of modelling for the infrared- and collinear-safe angularities is sensitive to the quark and gluon composition of the sample of jets.
Although grooming is expected to reduce the influence from pileup, underlying event, and initial-state radiation, which are all difficult to model, we find no significant improvement in the description with grooming.
A calculation based on analytic resummation of large logarithms of the collinear-safe angularities in the \Zjet topology at next-to-leading order + next-to-leading logarithm (NLO+NLL') accuracy including nonperturbative effects best describes the thrust with a $\chi^2/N_{\text{bins}}$ (where $N_{\text{bins}}$ is the number of bins of the distribution) as low as $4.1/5.0$, whereas the Les Houches Angularity was described significantly worse with $\chi^2/N_{\text{bins}}$ up to $58/8$.
A comparison of the means of the angularities in quark- and gluon-enriched data samples demonstrated their discrimination power, which was overestimated by all generators under study, showing a clear need for improvements in the simulation.

\begin{acknowledgments}
    We congratulate our colleagues in the CERN accelerator departments for the excellent performance of the LHC and thank the technical and administrative staffs at CERN and at other CMS institutes for their contributions to the success of the CMS effort. In addition, we gratefully acknowledge the computing centres and personnel of the Worldwide LHC Computing Grid and other centres for delivering so effectively the computing infrastructure essential to our analyses. Finally, we acknowledge the enduring support for the construction and operation of the LHC, the CMS detector, and the supporting computing infrastructure provided by the following funding agencies: BMBWF and FWF (Austria); FNRS and FWO (Belgium); CNPq, CAPES, FAPERJ, FAPERGS, and FAPESP (Brazil); MES and BNSF (Bulgaria); CERN; CAS, MoST, and NSFC (China); MINCIENCIAS (Colombia); MSES and CSF (Croatia); RIF (Cyprus); SENESCYT (Ecuador); MoER, ERC PUT and ERDF (Estonia); Academy of Finland, MEC, and HIP (Finland); CEA and CNRS/IN2P3 (France); BMBF, DFG, and HGF (Germany); GSRI (Greece); NKFIA (Hungary); DAE and DST (India); IPM (Iran); SFI (Ireland); INFN (Italy); MSIP and NRF (Republic of Korea); MES (Latvia); LAS (Lithuania); MOE and UM (Malaysia); BUAP, CINVESTAV, CONACYT, LNS, SEP, and UASLP-FAI (Mexico); MOS (Montenegro); MBIE (New Zealand); PAEC (Pakistan); MSHE and NSC (Poland); FCT (Portugal); JINR (Dubna); MON, RosAtom, RAS, RFBR, and NRC KI (Russia); MESTD (Serbia); SEIDI, CPAN, PCTI, and FEDER (Spain); MOSTR (Sri Lanka); Swiss Funding Agencies (Switzerland); MST (Taipei); ThEPCenter, IPST, STAR, and NSTDA (Thailand); TUBITAK and TAEK (Turkey); NASU (Ukraine); STFC (United Kingdom); DOE and NSF (USA).
    
    \hyphenation{Rachada-pisek} Individuals have received support from the Marie-Curie programme and the European Research Council and Horizon 2020 Grant, contract Nos.\ 675440, 724704, 752730, 758316, 765710, 824093, 884104, and COST Action CA16108 (European Union); the Leventis Foundation; the Alfred P.\ Sloan Foundation; the Alexander von Humboldt Foundation; the Belgian Federal Science Policy Office; the Fonds pour la Formation \`a la Recherche dans l'Industrie et dans l'Agriculture (FRIA-Belgium); the Agentschap voor Innovatie door Wetenschap en Technologie (IWT-Belgium); the F.R.S.-FNRS and FWO (Belgium) under the ``Excellence of Science -- EOS" -- be.h project n.\ 30820817; the Beijing Municipal Science \& Technology Commission, No. Z191100007219010; the Ministry of Education, Youth and Sports (MEYS) of the Czech Republic; the Deutsche Forschungsgemeinschaft (DFG), under Germany's Excellence Strategy -- EXC 2121 ``Quantum Universe" -- 390833306, and under project number 400140256 - GRK2497; the Lend\"ulet (``Momentum") Programme and the J\'anos Bolyai Research Scholarship of the Hungarian Academy of Sciences, the New National Excellence Program \'UNKP, the NKFIA research grants 123842, 123959, 124845, 124850, 125105, 128713, 128786, and 129058 (Hungary); the Council of Science and Industrial Research, India; the Latvian Council of Science; the Ministry of Science and Higher Education and the National Science Center, contracts Opus 2014/15/B/ST2/03998 and 2015/19/B/ST2/02861 (Poland); the Funda\c{c}\~ao para a Ci\^encia e a Tecnologia, grant CEECIND/01334/2018 (Portugal); the National Priorities Research Program by Qatar National Research Fund; the Ministry of Science and Higher Education, project no. 14.W03.31.0026 (Russia); the Programa Estatal de Fomento de la Investigaci{\'o}n Cient{\'i}fica y T{\'e}cnica de Excelencia Mar\'{\i}a de Maeztu, grant MDM-2015-0509 and the Programa Severo Ochoa del Principado de Asturias; the Stavros Niarchos Foundation (Greece); the Rachadapisek Sompot Fund for Postdoctoral Fellowship, Chulalongkorn University and the Chulalongkorn Academic into Its 2nd Century Project Advancement Project (Thailand); the Kavli Foundation; the Nvidia Corporation; the SuperMicro Corporation; the Welch Foundation, contract C-1845; and the Weston Havens Foundation (USA).
\end{acknowledgments}

\bibliography{auto_generated}
\cleardoublepage \appendix\section{The CMS Collaboration \label{app:collab}}\begin{sloppypar}\hyphenpenalty=5000\widowpenalty=500\clubpenalty=5000\vskip\cmsinstskip
\textbf{Yerevan Physics Institute, Yerevan, Armenia}\\*[0pt]
A.~Tumasyan
\vskip\cmsinstskip
\textbf{Institut f\"{u}r Hochenergiephysik, Vienna, Austria}\\*[0pt]
W.~Adam, J.W.~Andrejkovic, T.~Bergauer, S.~Chatterjee, M.~Dragicevic, A.~Escalante~Del~Valle, R.~Fr\"{u}hwirth\cmsAuthorMark{1}, M.~Jeitler\cmsAuthorMark{1}, N.~Krammer, L.~Lechner, D.~Liko, I.~Mikulec, P.~Paulitsch, F.M.~Pitters, J.~Schieck\cmsAuthorMark{1}, R.~Sch\"{o}fbeck, M.~Spanring, S.~Templ, W.~Waltenberger, C.-E.~Wulz\cmsAuthorMark{1}
% ...
\vskip\cmsinstskip
% ...
1:  Also at TU Wien, Wien, Austria\\
% ...
\end{sloppypar}
\end{document}